\documentclass[journal=jpcbfk,manuscript=article]{achemso}
\usepackage[utf8]{inputenc}

\usepackage{longtable}
\usepackage{multirow}
\usepackage{amsmath}
\usepackage[usenames,dvipsnames]{color}
\usepackage{soul}
\usepackage{threeparttable}
\usepackage{xr}
\usepackage{graphicx} \usepackage{amsmath,amssymb}
\usepackage{caption} \usepackage{color} \usepackage{xcolor}
\usepackage{dcolumn} \usepackage{bm} \usepackage{float}
\usepackage{subcaption} \usepackage{chemformula}
\usepackage{textcomp} \usepackage{url} \usepackage[normalem]{ulem}
\usepackage{comment}

 \usepackage{microtype}
\usepackage{etoolbox}

\usepackage[unicode=true, bookmarks=true, bookmarksnumbered=true,
  bookmarksopen=true, bookmarksopenlevel=2, breaklinks=false,
  pdfborder={0 0 1}, backref=false, colorlinks=true, hidelinks
]{hyperref}

\SectionNumbersOn

\author{Cangtao Yin} \affiliation[University of Basel]{Department of
  Chemistry, University of Basel, Klingelbergstrasse 80, CH-4056
  Basel, Switzerland.}
  
\author{Markus Meuwly} \affiliation[University of Basel]{Department of
  Chemistry, University of Basel, Klingelbergstrasse 80, CH-4056
  Basel, Switzerland.} \email{m.meuwly@unibas.ch}
  
\title{Full Reaction Pathway Dynamics for Atmospheric Decomposition
  Reactions: The Photodissociation of H$_2$COO}

\begin{document}
\date{\today}

\begin{abstract}
Branching ratios for fragmentation channels of important meta- and
unstable species are essential for a molecular-level characterization
of atmospheric chemistry. Here, the molecular product channels for the
decomposition dynamics of the smallest Criegee intermediate, H$_2$COO,
are quantitatively investigated. Using a high-quality,
full-dimensional machine learned potential energy surface
(CASPT2/aug-cc-pVTZ), the translational, rotational, and vibrational
energy distributions of the CO$_2$+H$_2$, H$_2$O+CO, and HCO+OH
fragmentation channels were analyzed to elucidate partitioning of the
available energy. The CO$_2$ + H$_2$ product forms through two
different pathways that bifurcate after formation of the OCH$_2$O
intermediate. Along the direct pathway, CO$_2$ is preferentially
vibrationally excited with H$_2$in its vibrational ground state,
whereas for the indirect pathway going through formic acid, H$_2$ can
populate levels with $v > 0$. For all product channels passing through
energized formic acid, the lifetime distributions are described by
stretched exponentials with $\beta$ ranging from 1.1 to 1.7. This is a
clear signature of non-RRKM effects and suggests that the explicit
molecular dynamics needs to be followed for a quantitative and
realistic description of the photodissociation dynamics.
\end{abstract}


\noindent
Criegee intermediates\cite{criegee1949ozonisierung}, formed through
ozonolysis of alkenes, are central species for the chemical evolution
of Earth’s atmosphere. Their formation route
R$_1$R$_2$C=CR$_3$R$_4$+O$_3$ $\rightarrow$
R$_1$R$_2$COO+R$_3$R$_4$C=O implies that they are generated with
considerable, but largely unknown amounts of internal energy. Previous
calculations on the O$_3$+C$_2$H$_4$ cycloaddition reaction report a
gain of $-48.1$ kcal/mol of the H$_2$COO+CH$_2$O product relative to
the reactants.\cite{stanton:2015} CIs are chemically unstable. They
either decay through unimolecular
(photo)dissociation\cite{green2017selective,yin2017does,vereecken2017unimolecular}
or react with other compounds including SO$_2$, NO$_2$, or H$_2$O in
the
atmosphere.\cite{stone2014kinetics,qiu2019detection,yin2024revealing,chao2015direct,yin2018effect}
Their relevance for atmospheric chemistry lies in providing a fast,
non-photolytic conversion of ethene and ozone into stable end
products, including a prompt source of molecular hydrogen
H$_2$.\cite{francisco:2012} Although the lifetime of the smallest
Criegee intermediate, H$_2$COO, is rather short (seconds to
sub-seconds, depending on humidity),\cite{berndt:2012,welz:2012} the
global ubiquity of ethene ozonolysis implies that H$_2$COO contributes
measurably to the tropospheric H$_2$ budget and influences the closure
of carbon oxidation pathways.\cite{taatjes:2013} Furthermore, H$_2$COO
serves as a benchmark system for characterizing energy disposal and
unimolecular reaction dynamics in Criegee intermediates more
generally. Consequently, a quantitative characterization of the
molecular decomposition pathways of H$_2$COO is of great general
interest.\\

\noindent
Compared with H$_2$COO, the unimolecular decay of the next-larger CI,
{\it syn-}CH$_3$CHOO, has been studied extensively in both
experiments\cite{fang:2016,fang:2016deep} and
theory.\cite{MM.criegee:2021,MM.criegee:2023} This is partly because
the rate of OH formation from {\it syn-}CH$_3$CHOO is much higher than
for H$_2$COO, which facilitates its laboratory experimental
detection. Analysis of the product translational energy release for
{\it syn-}CH$_3$CHOO revealed\cite{MM.criegee:2023} that a
quantitative interpretation of the experimental observations requires
tracking the full reaction evolution, from the initial non-equilibrium
preparation of the reactant through intermediates to the final
products.\cite{lester:2016} Starting reactive molecular dynamics (MD)
simulations from intermediates or transition states along the
decomposition pathway provided qualitative but not quantitative
information about the final state distributions.\cite{lester:2016}\\

\noindent
The present work aims at characterizing in a quantitative fashion the
end-to-end reaction dynamics of H$_2$COO, with a particular focus on
the analysis of final product states and bifurcating dynamics. To
achieve this, additional reference data in the product regions was
included into the training of the previously developed global
potential energy surface (PES).\cite{MM.h2coo:2025} This yields a
suitably extended and improved machine learning-based PES (ML-PES) for
final state analysis based on CASPT2 reference data, which provides an
accurate and unified description of all possible reaction
channels. The improved coverage of the product regions ensures a
reliable representation of bond breaking and formation processes and
allows long-time trajectory propagation. \\

\begin{figure}[h!]
    \centering \includegraphics[width=0.9\linewidth]{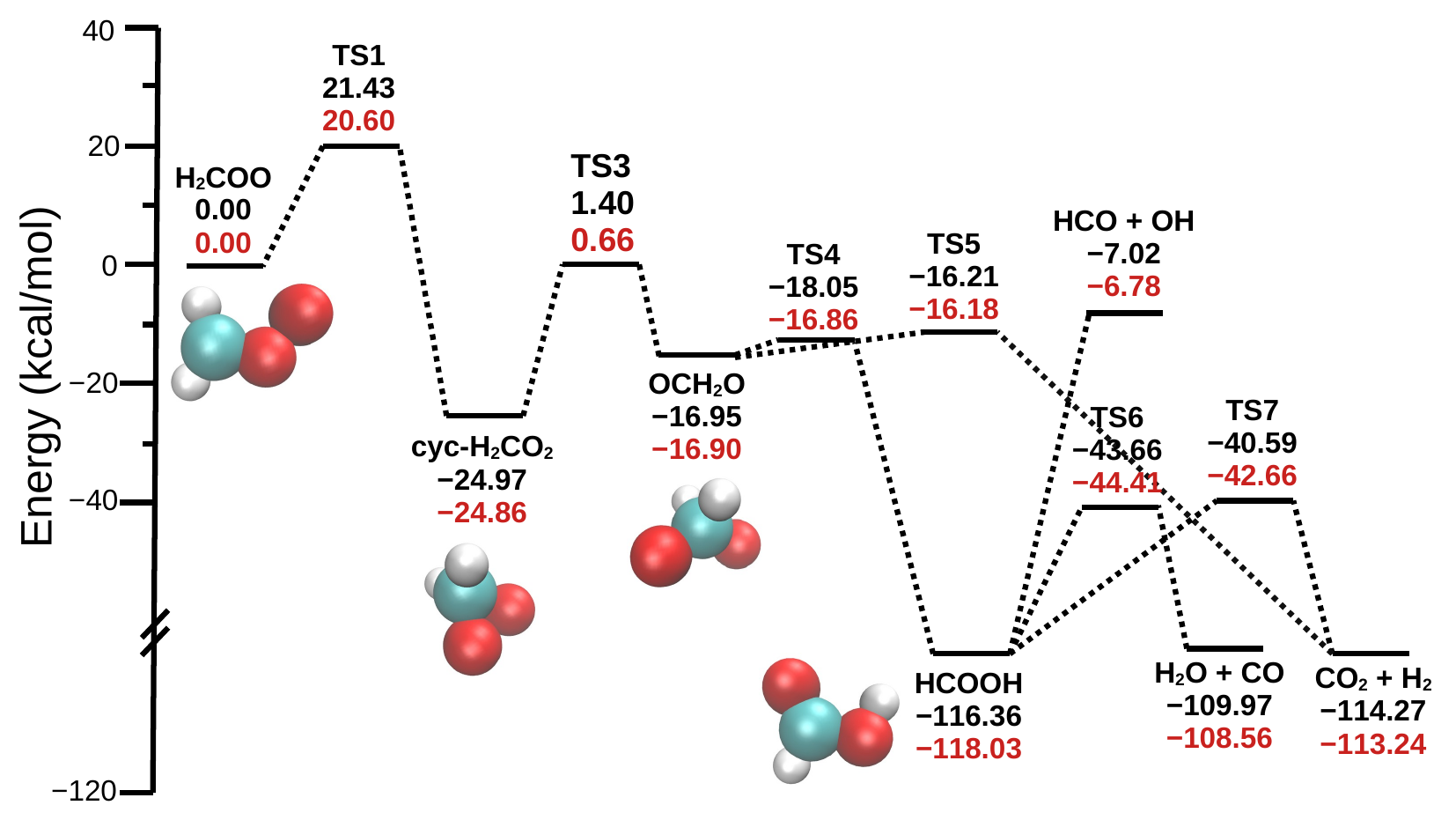}
    \caption{The unimolecular decomposition of H$_2$COO into three
      bimolecular products: CO$_2$ + H$_2$, H$_2$O + CO, and HCO +
      OH. The formation of CO$_2$ + H$_2$ proceeds through two
      competing pathways: the ``direct pathway'', which connects
      OCH$_2$O directly to the products via TS5, and the ``indirect
      pathway'', in which OCH$_2$O first isomerizes to HCOOH (formic
      acid) before forming CO$_2$ + H$_2$.}
    \label{fig:diagram}
\end{figure}

\noindent
The performance of the trained ML-PESs was evaluated by comparing
NN-predicted energies $E$ and forces $F$ with reference CASPT2
calculations, see Figure \ref{sifig:corr}. Specifically, the mean
absolute (MAE) and root mean squared errors (RMSE) were computed for
both quantities using a test set of approximately 1500 structures. For
the best-performing PhysNet model, the MAE$(E)$ and RMSE$(E)$ are 1.23
kcal/mol and 2.14 kcal/mol, respectively. The corresponding errors for
forces are ${\rm MAE}(F) = 1.11$ kcal/mol/\AA\/ and ${\rm RMSE}(F) =
3.80$ kcal/mol/\AA\/. Remarkably, the data set spans $\sim 150$
kcal/mol, as shown in Figure \ref{fig:diagram}. Furthermore, the model
shows a very high correlation with the reference data, with $R^2 =
0.9984$, demonstrating that it accurately reproduces the trends and
variations in the CASPT2 energies. The performance of the present
model is comparable in accuracy to previously reported
results,\cite{MM.h2coo:2025,MM.h2coo:2024,MM.criegee:2023}, indicating
that the ML-PES maintains the reliability of earlier models while
extending their applicability to the dissociation products.\\

\noindent
For the three molecular product channels, 1- and 2-dimensional PESs
are reported in Figures \ref{sifig:diatom} to
\ref{sifig:cont_hco}. For the product diatomics H$_2$, CO, and OH, all
1-d curves are globally smooth and well-behaved which is notable
because the full-dimensional PES is trained as a whole. Also, the
energy profiles only depend moderately on the separation $d$ away from
the respective triatomic interaction partner. At large diatom--triatom
separations ($d \sim 5$ \AA\/, blue traces in Figure
\ref{sifig:diatom}), the intermolecular interactions decrease to a
degree that the energy profiles approach those of the isolated di- and
triatomic species. In the asymptotic limit, a crossing appears for the
HCO + OH channel. However, this feature occurs at energies well above
the initial energy of the simulations (approximately 25.5 kcal/mol)
and therefore does not influence the MD trajectories, considering that
the initial energy of simulation is around 25.5 kcal/mol.\\

\noindent
Figures \ref{sifig:cont_co2} to \ref{sifig:cont_hco} illustrate the
PESs of the triatomic products CO$_2$, H$_2$O, and HCO in the presence
of their corresponding diatomic partners, H$_2$, CO, and OH,
respectively. Again, all PESs are smooth and well-behaved which allow
to run valid reactive MD simulations and to analyze the vibrational
contribution to the total internal energy of the fragments. For CO$_2$
and short intramolecular separation ($d = 1.70$ \AA\/, see Figure
\ref{sifig:cont_co2}A) the energy profile of two CO bond stretching is
not symmetric with respect to the diagonal, since the interaction
between CO$_2$ and H$_2$ breaks this symmetry. With increasing $d$ the
expected symmetry of $V(r_{\rm CO_A},r_{\rm CO_B})$ becomes
apparent. The C--O dissociation energy for the oxygen atom away from
CO$_2$ from the NN-PES is $\sim 130$ kcal/mol which agrees favouarbly
with the experimentally reported value of 126
kcal/mol.\cite{gong2024bond} This supports the reliability of the
present PES. For CO$_2$ + H$_2$, the CO bonds remain intact because
dissociation requires much higher energy and in the MD trajectories
some energy is carried away by H$_2$, making CO bond cleavage
unlikely. For the H$_2$O + CO system (see Figure \ref{sifig:cont_h2o},
variations in the OH bond lengths are similarly explored across
different H$_2$O–CO separations, revealing a stable triatomic product
region under the simulation conditions.\\

\noindent
For HCO + OH, the potential energy minimum rises with increasing
HCO–OH separation, consistent with the endothermic character of HCOOH
decomposition. For the HCO triatomic, the CH bond dissociation energy
from the present NN-PES is $\sim 15$ kcal/mol. This compares well with
earlier measurements which reported values of [13.36, 14.40, 14.54]
kcal/mol and validates this
3d-PES.\cite{peters:2013,vichietti:2020,sun:2024} In addition, along
the H-atom dissociation pathway, a transition state is observed in
this figure. This transition state was also identified in earlier
MRCI/aug-cc-pVQZ\cite{peters:2013} and
CCSD(T)/CBS//CCSD/aug-cc-pVTZ\cite{vichietti:2020} studies and lies
below the initial 25.5 kcal/mol of energy available in the MD
simulations. As a result, a small fraction of trajectories proceeded
to form the H + CO + OH products. The minimum rises with increasing
$d{\rm (HCO, OH)}$, in contrast to the CO$_2$+H$_2$ and H$_2$O+CO
cases. This behavior is expected, as HCOOH decomposition along this
pathway proceeds without a transition state.\\

\noindent
Following previous experimental\cite{lester:2016,lester:2024} and
computational\cite{MM.criegee:2021,MM.criegee:2023,MM.h2coo:2024,MM.h2coo:2025}
work, the reactant H$_2$COO was energized through selective
vibrational excitation (see Methods). Such an excitation scheme mimics
energy localization in specific internal modes and allows the system
to efficiently access the reactive regions of the PES. As a result,
the full unimolecular decay of H$_2$COO can be followed dynamically,
enabling a direct examination of the reactions and the associated
nonequilibrium dynamical effects governing product formation. In
particular, the translational, rotational and vibrational energies
characterizing each fragment can be analyzed and compared, see
Methods. Intermediate structures sampled along the reaction pathways
are identified through geometrical criteria summarized in Table
\ref{sitab:criteria}\\

\begin{figure}[h!]
    \centering \includegraphics[width=0.95\linewidth]{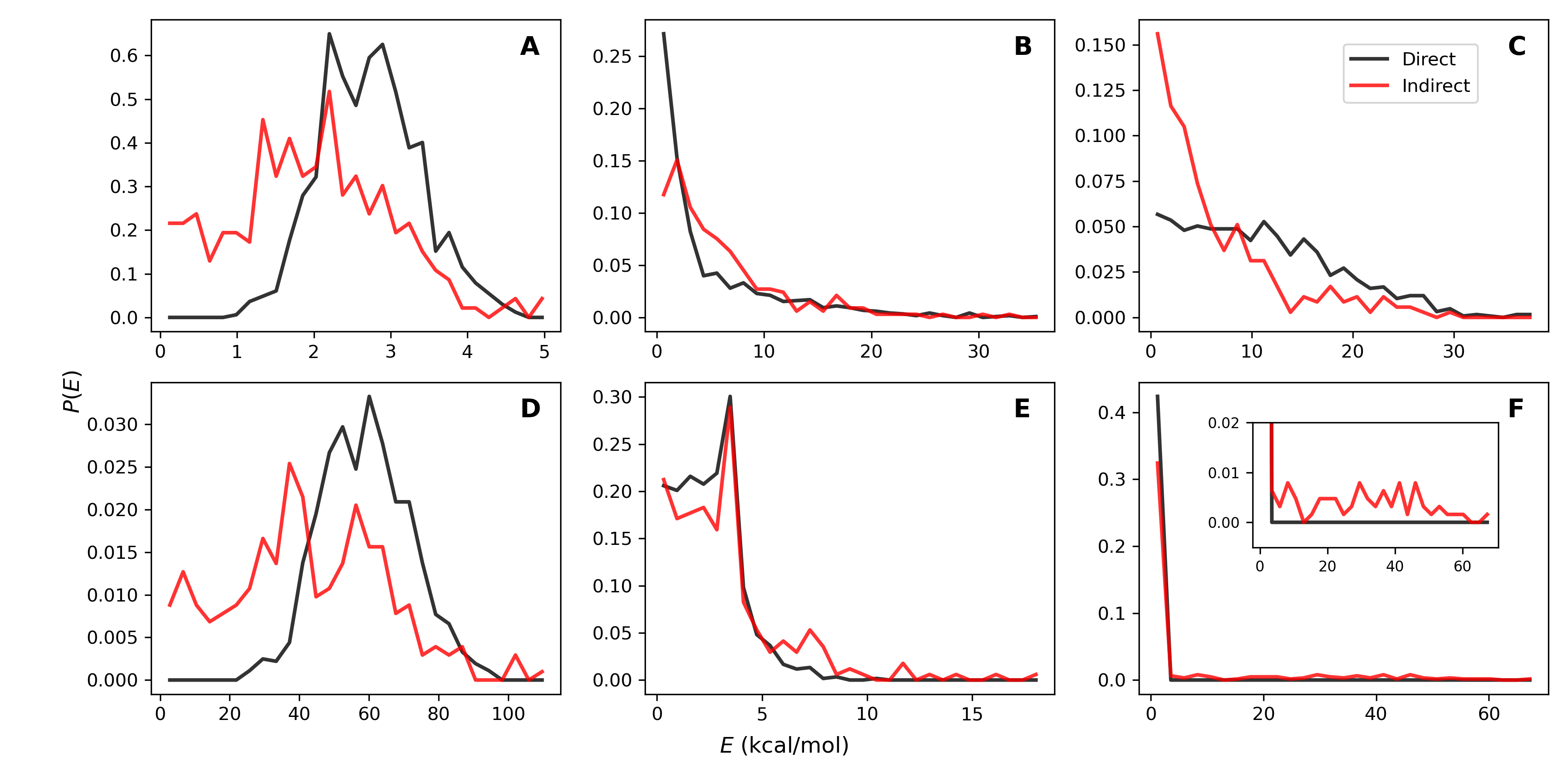}
    \caption{Energy distributions for the CO$_2$+H$_2$ product
      channel. Trajectories follow direct (952) and indirect (268)
      routes. Top panels: fragment CO$_2$; bottom panels: fragment
      H$_2$. From left to right: translational, rotational, and
      vibrational energy distributions. The inset in panel F indicates
      that $v > 0$ can be populated for H$_2$ along the indirect
      pathway. Note the different scales along the $x-$axes for panels
      A-C vs. D-F.}
    \label{fig:co2_h2_dir_ind}
\end{figure}

\noindent
{\bf Product State Analysis:} With validated product state potential
energy surfaces, a total of 4000 nonequilibrium ML-MD simulations was
run (see Methods): 952 and 268 formed CO$_2$ + H$_2$ directly and
indirectly through HCOOH, respectively, 664 formed H$_2$O + CO, and 40
formed HCO + OH. The remaining trajectories do not progress beyond the
reactant and decay on longer time scales. The energy distributions of
the final products were analyzed separately for each of the 4
identified pathways. Formation of CO$_2$ + H$_2$ (Figure
\ref{fig:co2_h2_dir_ind}) along the direct (black traces) pathway
yields Gaussian-shaped translational energy distributions for CO$_2$
(panel A) and H$_2$ (panel D), with H$_2$ acquiring particularly high
translational energies due to its low mass. For the indirect pathway
(red traces in panels A and D) the shapes of $P(E_{\rm trans})$
deviate from a Gaussian and it is noted that $E_{\rm trans}$ is
generally higher along the direct pathway compared to the indirect
pathway.\\

\noindent
The rotational energy distributions are comparable for both pathways,
see Figures \ref{fig:co2_h2_dir_ind}B/E. Evidently, $P(E_{\rm rot})$
for H$_2$ is not in thermal equilibrium, but given the large
rotational constant of H$_2$ ($B = 61$ cm$^{-1}$) only $j=0$ and
potentially $j=1$ are expected to be populated. In contrast, the
vibrational energy shows a distinct difference: CO$_2$ (panel C) gains
more energy along the direct pathway, whereas for H$_2$ (panel F) the
vibrational degree of freedom contains sufficient energy along the
indirect pathway to excite states with $v > 0$ given that the
fundamental H$_2 (v=1)$ appears at 4400 cm$^{-1}$, corresponding to
$\sim 12.6$ kcal/mol. This is also an experimentally accessible
quantity which allows to distinguish the two pathways.\\

\begin{figure}[h!]
    \centering \includegraphics[width=0.95\linewidth]{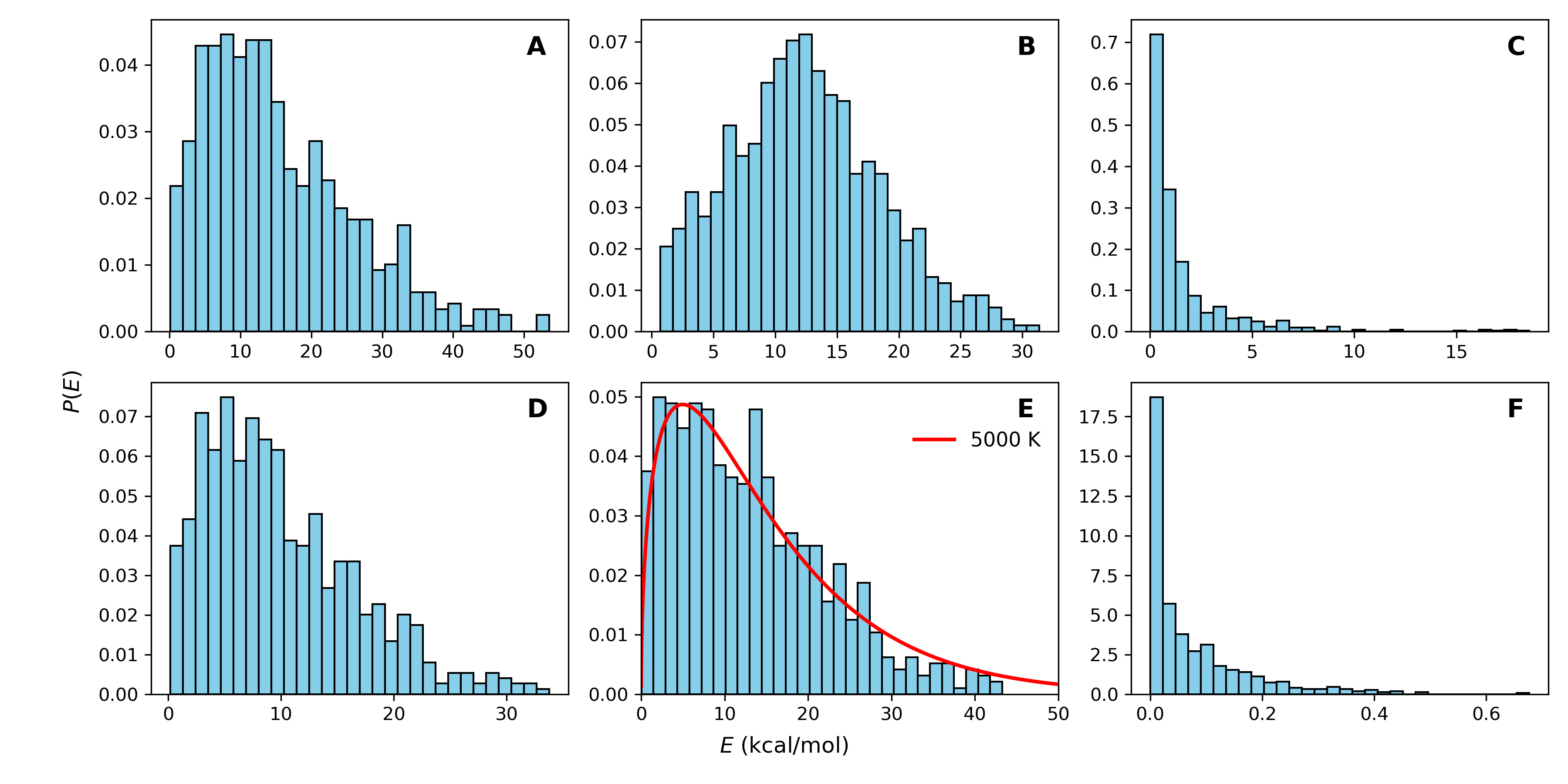}
    \caption{Energy distributions of the product H$_2$O+CO from 664
      simulations. The top panels show fragment H$_2$O and bottom show
      fragment CO. From left to right, the distributions correspond to
      translational, rotational, and vibrational energies. The red
      trace in panel E corresponds to a Boltzmann distribution with $T
      = 5000$ and suggests that the rotational motion of dissociating
      CO is close to thermal equilibrium.}
    \label{fig:h2o_co_his}
\end{figure}

\noindent
For the H$_2$O + CO channel, the energy partitioning into
translational, rotational, and vibrational contributions is shown in
Figure \ref{fig:h2o_co_his}. The CO product remains exclusively in the
vibrational ground state (panel F) as $\sim 6$ kcal/mol (2143
cm$^{-1}$) is required to reach $v = 1$. For H$_2$O, on the other
hand, $P(v)$ extends out to $\sim 20$ kcal/mol which suffices to
populate vibrational levels: the 3 fundamental vibrations (1595, and
3657 to 3756 cm$^{-1}$) require 4.6 and 10.5 to 11.0 kcal/mol for
populating the bending and stretching modes, respectively. For CO, the
Boltzmann distribution $P(j) = \exp[-\beta E_{\rm rot}(j)]$ for
$T=5000$ K and with $E_{\rm rot} = B j (j + 1)$ where $B = 1.93~{\rm
  cm}^{-1}$ closely follows the histogram which indicates that the
product diatomic is close to rotational equilibrium. However, there is
also a “plateau” in $P(E_{\rm rot}^{\rm CO})$ at low rotational
energies which is reminiscent of that seen for OH-dissociation from
vibrationally excited {\it syn-}CH$_3$CHOO. In that case a plateau
spanning rotational quantum numbers from 3 to 6 was observed in both,
experiment\cite{lester:2016} and simulations.\cite{MM.criegee:2023}\\

\noindent
Finally, for HCO + OH (Figure \ref{sifig:hco_oh_his}), the final state
distributions are not sufficiently well-converged because only 40
trajectories contribute to this fragmentation channel. Nevertheless,
it is evident, that the translational energies are considerably
smaller ($E_{\rm trans}^{\rm HCO} \lesssim 7$ kcal/mol and $E_{\rm
  trans}^{\rm OH} \lesssim 10$ kcal/mol) than for the other two
pathways. The vibrational energies ($E_{\rm vib}^{\rm HCO} \lesssim 6$
kcal/mol and $E_{\rm vib}^{\rm OH} \lesssim 0.025$ kcal/mol) only
suffice to excite the HCO bending and CO local modes (1101 cm$^{-1}$
and 1870 cm$^{-1}$, equivalent to 3.1 kcal/mol and 5.3 kcal/mol), but
not the CH and OH stretch modes (2750 and 3737 cm$^{-1}$,
corresponding to 7.9 and 10.7 kcal/mol). The only notable excitations
are for the rotational degrees of freedom, for which $\sim 10$
kcal/mol are available for both reaction products.\\

\noindent
{\bf Bifurcating Pathway to yield CO$_2$ + H$_2$:} As noted earlier,
formation of CO$_2$ + H$_2$ can follow two competing pathways. With an
initial energy content of 25.5 kcal/mol in the reactant (H$_2$COO),
the excess energy in OCH$_2$O is $> 40$ kcal/mol. Hence, OCH$_2$O is
unstable (lifetime $< 0.1$ ps) and the reaction pathway towards
CO$_2$+H$_2$ bifurcates into a direct and an indirect route through a
formic acid (HCOOH) intermediate, see Figure \ref{fig:diagram}.\\

\begin{figure} [H]
    \centering
    \includegraphics[width=0.8\linewidth]{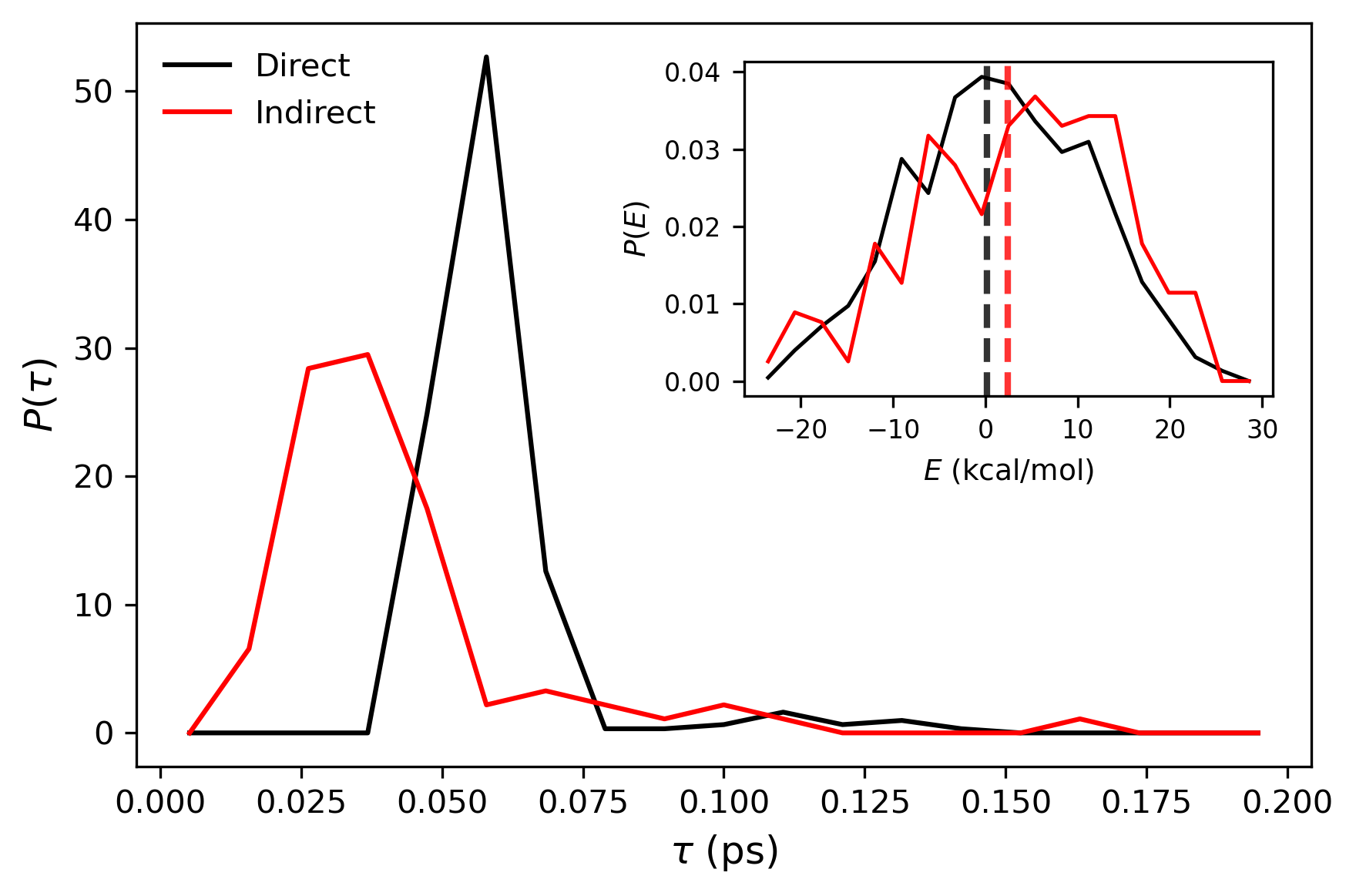}
    \caption{Lifetime (main view) and energy (inset) distributions of
      OCH$_2$O for the direct (black) and indirect (red) channels. In
      both cases, OCH$_2$O is short-lived, with lifetimes never
      exceeding 0.2 ps. In the inset, the median energy (dashed
      vertical lines) are at 0.2 and 2.4 kcal/mol, for the direct and
      indirect cases, respectively.}
    \label{fig:dioxy_time_e}
\end{figure}

\noindent
Notably, the OCH$_2$O lifetime distributions for the direct (black)
and indirect (red) pathways differ substantially, see Figure
\ref{fig:dioxy_time_e}. The distribution for the direct pathway is
shifted toward longer times, with a peak at approximately 60 fs,
whereas the indirect pathway exhibits a maximum at $\sim 30$
fs. Indeed, the configurational energy distributions of the OCH$_2$O
intermediate from trajectories following the direct (black) and
indirect (red) pathways reflect the differences in the lifetimes:
along the direct pathway the maximum of $P(E_{\rm int})$ (centered
around 0) is shifted towards lower energies than for the indirect
pathway (peak at 5 kcal/mol) which extends the lifetime for this
intermediate and their median values (dashed lines) differ by 2.2
kcal/mol. Other than that, the two energy distributions are comparable
in terms of their shape.\\

\begin{figure} [H]
    \centering \includegraphics[width=1.0\linewidth]{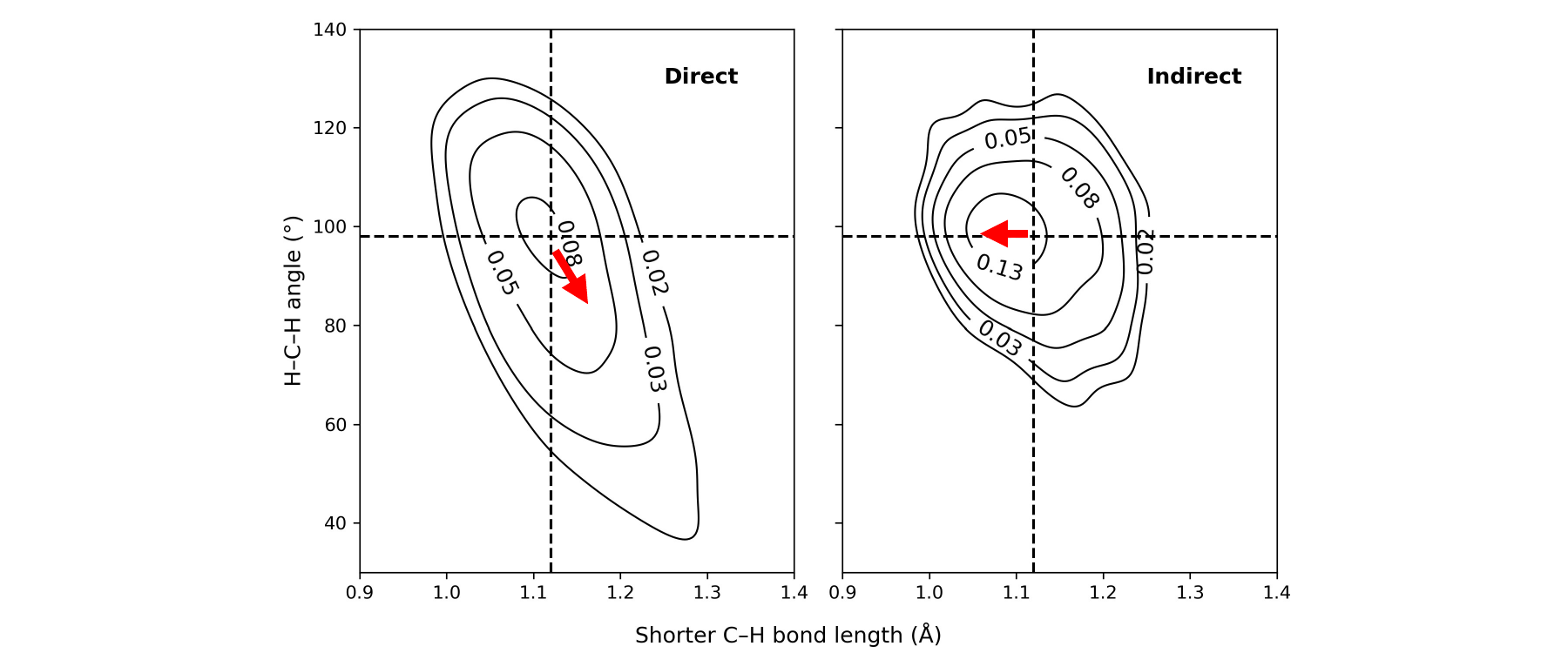}
    \caption{Normalized probability distributions $P(r_{\rm
        CH},\theta_{\rm HCH})$ while sampling the OCH$_2$O well for
      each of the direct (left) and indirect (right) channels. The
      numbers in the contour lines indicate the normalized intensity
      of geometries. The dashed lines represent the equilibrium
      structure of OCH$_2$O, and the geometrical criteria used to
      assign sampled structures to each product channel are listed in
      Table \ref{sitab:criteria}. The red arrows illustrate the
      reaction pathways from OCH$_2$O to the corresponding reaction
      pathways.}
    \label{fig:dioxy_ch_hch}
\end{figure}

\noindent
The differences between the two pathways become even more evident when
analyzing structural properties. Figure \ref{fig:dioxy_ch_hch} reports
the $P(r_{\rm CH},\theta_{\rm HCH})$ distribution relating the shorter
of the two CH bond lengths with the H-C-H bond angle in OCH$_2$O. For
the direct pathway, the H–C–H angle distribution extends to
significantly smaller values ($40^{\circ}$), and the CH bond length
stretches considerably more away from the equilibrium structure. These
two motions are required for (direct) H$_2$ formation. In contrast,
the indirect pathway exhibits a C–H bond length distribution shifted
toward shorter values. This trend is consistent with the fact that the
C–H bond length in HCOOH (1.09 \AA\/) is shorter than that in OCH$_2$O
(1.12 \AA\/). Together, these differences in the joint geometric
distributions highlight that the reaction outcome is closely linked to
the instantaneous molecular geometry, reinforcing the picture that
molecular motion rather than total energy alone, plays a key role in
steering the reaction pathways.\\

\noindent
{\bf Breakup Dynamics of Unstable HCOOH:} With an excess energy of
$\sim 140$ kcal/mol, HCOOH formed through nonequilibrium preparation
of the reactant is highly unstable, in particular with respect to the
two lower-lying product channels CO$_2$+H$_2$ and H$_2$O+CO, see
Figure \ref{fig:diagram}. The energy distributions of HCOOH associated
with the three product channels, HCO+OH, H$_2$O+CO, and CO$_2$+H$_2$,
are similar, see Figure \ref{sifig:fa_e}. This behavior serves as a
useful sanity check, as no clear distinctions are observed in either
the overall shape or the accessible energy range. It confirms that the
total internal energy of HCOOH alone does not play a decisive role in
determining the subsequent reaction outcome. This suggests that the
branching into different product channels is not governed by energy
selection alone, but instead arises from dynamical effects such as
mode specific energy flow and structure evolution. The similarity of
the energy distributions further supports the conclusion that HCOOH
dissociates under strongly nonequilibrium conditions, where the
instantaneous configuration and dynamics of the intermediate, rather
than its total energy, control reaction pathway selection.\\

\begin{figure}[h!]
    \centering \includegraphics[width=1.0\linewidth]{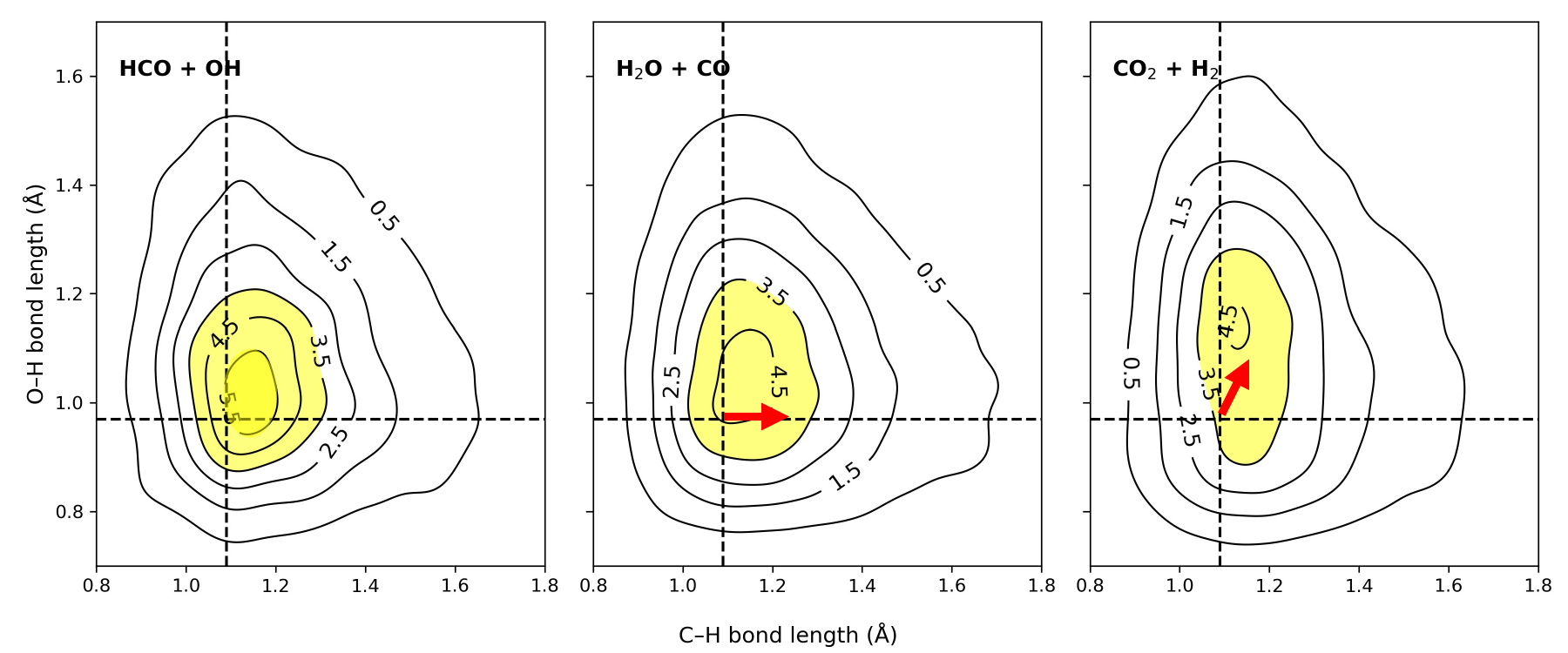}
    \caption{Normalized probability distributions $P(r_{\rm CH},
      r_{\rm OH})$ sampled in the HCOOH well for the three product
      channels: HCO + OH (left), H$_2$O + CO (middle), and CO$_2$ +
      H$_2$ (right). The dashed lines indicate the equilibrium
      geometry of HCOOH, and the geometrical criteria used to assign
      sampled structures to each product channel are listed in Table
      \ref{sitab:criteria}. The numbers on the contours denote the
      normalized population density. Red arrows in the middle and
      right panels illustrate the reaction pathways from HCOOH to the
      corresponding products.}
    \label{fig:fa_ch_oh}
\end{figure}

\noindent
Figure \ref{fig:fa_ch_oh} shows the joint distributions $P(r_{\rm
  CH},r_{\rm OH})$ relating the OH and CH bond lengths in HCOOH. For
trajectories leading to HCO+OH (left panel), the distribution is
strongly localized near the equilibrium geometry of formic acid, with
CH and OH bond lengths close to 1.09 \AA\/ and 0.97 \AA\/,
respectively, which are close to the equilibrium bond lengths in the
HCO and OH fragments. This allows dissociation to occur without
substantial distortion of the HCO and OH subunits within the HCOOH
molecule. In contrast, trajectories that produce CO$_2$ + H$_2$ (right
panel) exhibit systematically elongated O–H bond lengths: the maximum
of $P(r_{\rm CH},r_{\rm OH})$ is at $r_{\rm OH} = 1.15$ \AA\/,
considerably away from the equilibrium structure (dashed horizontal
line). This reflects the progressive cleavage of the OH bond along
this reaction pathway. The broader spread of geometries in this region
highlights the role of specific bond elongation in steering the system
toward different products. Compared to the CO$_2$ + H$_2$ channel
(right panel), the H$_2$O + CO channel (middle panel) exhibits
pronounced C–H bond stretching but no O–H bond stretching. This
behavior is expected, as formation of H$_2$O + CO requires cleavage of
the C–H bond, whereas the O–H bond remains intact, as indicated by the
red arrows illustrating the reaction directions. Together, these
geometric distributions demonstrate that product branching is closely
tied to the instantaneous molecular structure of HCOOH rather than to
its total internal energy, reinforcing the picture of a dynamically
controlled bifurcation.\\

\noindent
The three product channels also differ in the H--H separations that
are sampled when leaving the distribution of FA-geometries. Figure
\ref{sifig:HH_distances} shows the time evolution of the H–H
separation along the three pathways for five representative reactive
MD trajectories each. The zero of time is shifted so that $t=0$
corresponds to the frame at which products start to be formed (see
Table \ref{sitab:criteria}).  While still in the well of HCOOH, the
H–H distance fluctuates around 2.79 \AA\/, consistent with the
equilibrium H–H separation in HCOOH. As the reaction proceeds,
distinct behaviors emerge for each channel. For trajectories forming
HCO+OH (black), the H–H distance increases sharply near $t=0$,
reflecting the separation of the two fragments. In contrast,
trajectories yielding H$_2$O+CO (red) show stabilization of the H–H
distance around 1.52 \AA\/, characteristic of the equilibrium geometry
of water, whereas those producing CO$_2$+H$_2$ (blue) exhibit
fluctuations around 0.74 \AA\/, corresponding to the H–H equilibrium
bond length in H$_2$.\\

\begin{figure}[h!]
    \centering \includegraphics[width=0.8\linewidth]{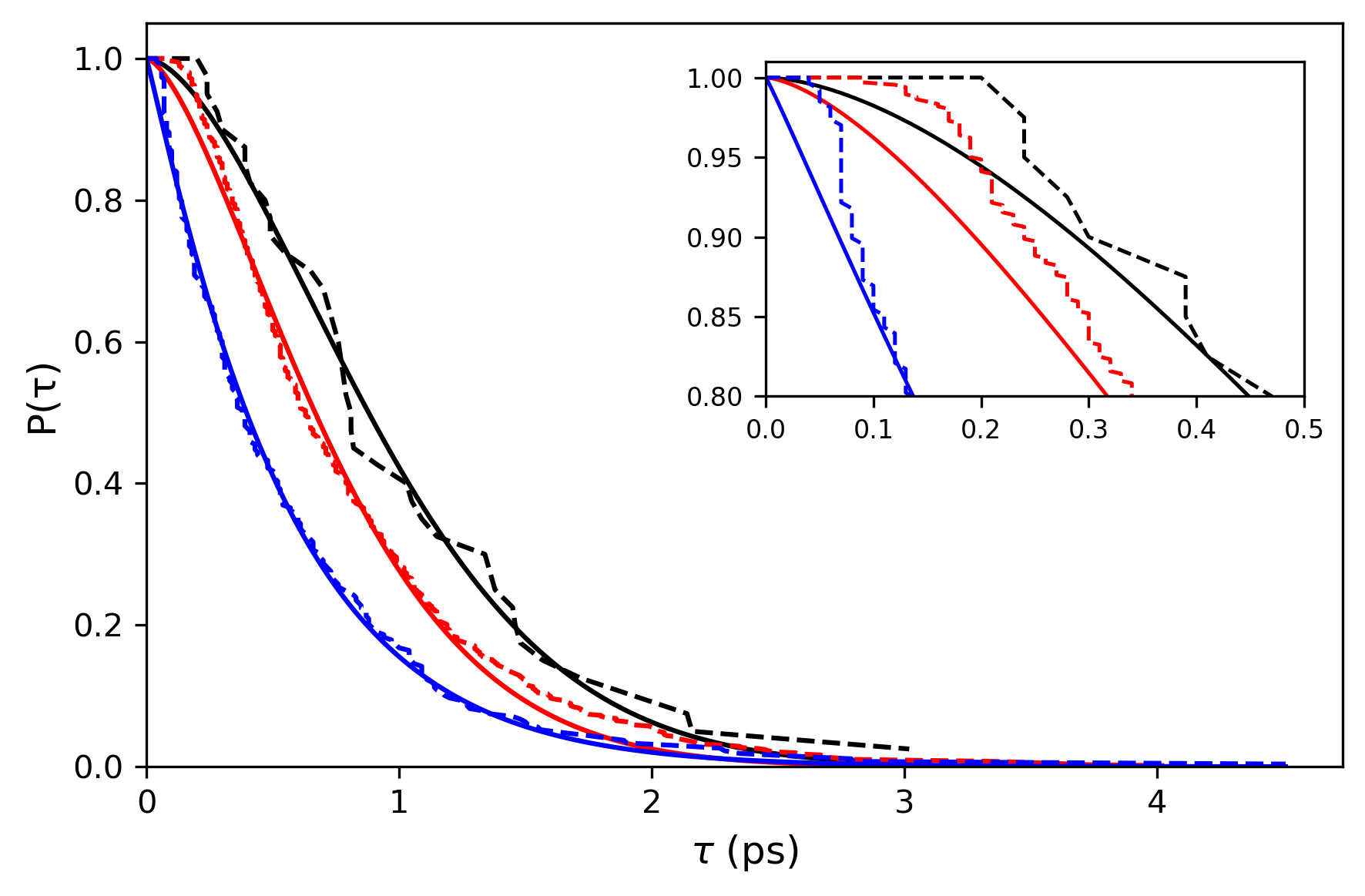}
    \caption{Lifetime distributions of HCOOH associated with the three
      product channels HCO+OH (black), H$_2$O+CO (red), and
      CO$_2$+H$_2$ (blue). The dashed lines represent the raw data,
      while solid lines show the fits using a stretched exponential
      function, $P(\tau) = \exp\left[-(\tau / \tau_{\rm
          c})^{\beta}\right]$, with $\tau_{\rm c} = 1.1; 0.8; 0.6$ ps
      and $\beta = 1.7; 1.5; 1.1$ for the HCO+OH; H$_2$O+CO;
      CO$_2$+H$_2$ pathways, respectively. The inset shows a magnified
      view of the early-time decay of the survival
      probability. Notably, for all three pathways there is a finite
      lifetime ranging from 75 fs to 200 fs during which no product is
      formed. Depending on the product channel considered, this
      HCOOH-lifetime differs by a factor of 3.}
    \label{fig:fa_time}
\end{figure}

\noindent
The most profound difference between the three reaction channels
emerging from energized HCOOH is found for the lifetime distribution
$P(\tau)$ of HCOOH before breakup, see Figure \ref{fig:fa_time}. To
characterize these distributions quantitatively, they were fitted to
stretched exponential functions $P(\tau) = e^{(-\tau / \tau_{\rm
    c})^\beta}$, with $\tau$ the lifetime and $\tau_{\rm c}$ the
characteristic time scale on which $P(\tau)$ decays. The use of
stretched-exponential kinetics was originally motivated by empirical
observations that relaxation and decay processes could not be
adequately described by finite sums of exponentials over extended time
windows.\cite{kohlrausch:1854,watts:1970} Subsequently, it was
recognized that the stretched exponential provides a compact and
robust representation of relaxation arising from a broad, continuous
distribution of timescales or energy
barriers.\cite{austin:1974,austin:1975,richert:2002} While also
offering a parameter-efficient alternative to multi-exponential fits,
its primary value lies in encoding physical heterogeneity through the
stretch exponent $\beta$ rather than in minimizing the number of fit
parameters.\\

\noindent
The raw survival curves show a short-time plateau preceding the onset
of decay, which can be interpreted as an apparent initialization
time. During this early period, trajectories remain in the reactant
region and do not immediately lead to product formation. The extent of
this plateau varies among the reaction channels: the HCO+OH channel
(black) exhibits the longest delay, while the CO$_2$+H$_2$ channel
displays the shortest. Although the precise origin of this behavior
cannot be uniquely determined from the present data, the observed
channel dependence suggests that different preparatory intramolecular
motions or structural fluctuations may be required before the system
can access the corresponding dissociation pathways.\\

\noindent
In the present case, the fitted parameters $[\beta, \tau_{\rm c}]$
differ considerably for the three product channels. This is
particularly apparent for the stretch exponent $\beta$. A stretch
exponent $\beta \sim 1$ indicates near single-exponential decay,
consistent with statistical unimolecular kinetics in the RRKM limit,
whereas $\beta > 1$ points towards compressed-exponential behaviour
which is not associated with statistical unimolecular kinetics. On the
other hand, $0 < \beta < 1$ points towards dynamic heterogeneity,
traps, or rough energy landscapes as they appear for
glasses,\cite{bauchy:2015} for
proteins,\cite{wolynes:1991,MM.mbno:2016} or for diffusion on rough
surfaces such as amorphous solid water.\cite{MM.o2:2019} In the
present case, clearly all values $\beta > 1$. For CO$_2$+H$_2$, $\beta
= 1.1$ which is closest to RRKR-like dynamics whereas for the the
H$_2$O+CO and HCO+OH channels $\beta = 1.5$ and $\beta = 1.7$,
respectively, which is evidently non-RRKM. Remarkably, Figure
\ref{fig:dioxy_time_e} does not report any difference between the three
pathways, whereas the HCOOH lifetime distribution $P(\tau)$ allows to
distinguish them and provides deeper physical insights. One slight
caveat concerns the HCO+OH channel for which only 40 events are
available, whereas for the two lower-lying product channels
statistically significant numbers of trajectories have been
analyzed. Nevertheless, the conclusion from this analysis remains that
non-RRKM dynamics is found for all three channels.\\

\noindent
Overall, the present work found non-RRKM dynamics for all three
molecular fragmentation channels for the unimolecular decay of
vibrationally excited H$_2$COO. For the H$_2$-production channel,
$\beta = 1.7$ is largest which points towards pronounced non-RRKM
characteristics of this decay pathway going through formic acid. Also,
the product H$_2$ generated along this pathway is in thermal
nonequilibrium and can populate vibrationally excited states of the
diatomic, see red trace Figure \ref{fig:co2_h2_dir_ind}F. For the
direct route, H$_2$ remains in its vibrational ground state which
provides a way to distinguish the two pathways in experiments.\\

\noindent
Analysis of joint probability distributions of internal coordinates
clarifies their role in selecting a particular reaction pathways. For
the bifurcating reaction, $P(r_{\rm CH}, \theta_{\rm HCH})$ in Figure
\ref{fig:dioxy_ch_hch}, clearly distinguishes between the direct and
the indirect pathway for CO$_2$+H$_2$ formation.  Likewise, $P(r_{\rm
  CH}, r_{\rm OH})$ provides a molecular-level understanding for the
differences between the three product channels emanating from
energized HCOOH.\\

\noindent
In conclusion, this study provides a mechanistic and quantitative
picture for the photodissociation dynamics of energized H$_2$COO under
near-laboratory conditions.\cite{lester:2016,lester:2024} Together
with earlier work on {\it syn-}Criegee, it is found that statistically
significant numbers of trajectories on ML-trained PESs at the CASPT2
level of theory provide a meaningful starting point for
molecular-level exploration of these important and challenging
systems.\cite{MM.criegee:2023,MM.h2coo:2025} Concerning the important
role of H$_2$COO to contribute measurably to the tropospheric H$_2$
budget, it was explicitly shown that H$_2$ can be formed exclusively
in the ground vibrational state (direct) or populate vibrationally
excited states for the reaction along the indirect route through
formic acid. Finally, it is demonstrated, that non-RRKM dynamics is at
play which underlines the importance of dynamical simulations for
understanding complex, multichannel reactions that cannot be fully
described by kinetic studies alone.\\

\section*{Supporting Information} 
The supporting information describes the Methods (training of PhysNet,
MD simulations, energy decomposition analysis) and provides additional
information about the product state PESs together with figures and
tables.

\section*{Data Availability} 
The reference data that allow to reproduce the findings of this study
are openly available at  \\

\section*{Acknowledgment}
Financial support from the Swiss National Science Foundation through
grants $200020\_219779$ (MM), $200021\_215088$ (MM), the University of
Basel (MM) is gratefully acknowledged. This article is based upon work
within COST Action COSY CA21101, supported by COST (European
Cooperation in Science and Technology) (to MM).


\clearpage

\renewcommand{\thetable}{S\arabic{table}}
\renewcommand{\thefigure}{S\arabic{figure}}
\renewcommand{\thesection}{S\arabic{section}}
\renewcommand{\d}{\text{d}}
\setcounter{figure}{0}  
\setcounter{section}{0}  
\setcounter{table}{0}

\newpage

\noindent
{\bf SUPPORTING INFORMATION: Full Reaction Pathway Dynamics for
  Atmospheric Decomposition Reactions: The Photodissociation of
  H$_2$COO}

\section{Methods}
\subsection{Construction of ML-PES}
To achieve a faithful description of the full reactive landscape
connecting H$_2$COO to its bimolecular products CO$_2$+H$_2$,
H$_2$O+CO, and HCO+OH, the ML-PES employed in this study was refined
by augmenting an existing data set with additional reference points in
the product regions\cite{MM.h2coo:2025}. This extended data coverage
is a prerequisite and essential for an adequate description of the
asymptotic behavior of each product channel and for ensuring smooth
and stable trajectory propagation throughout the dissociation
processes in the ML-MD.\\

\noindent
All electronic structure calculations used to generate the reference
energies were performed at the CASPT2/aug-cc-pVTZ (CASPT2/aVTZ), using
the MOLPRO program package\cite{molpro:2020}. The underlying complete
active space self-consistent field (CASSCF) calculations were carried
out using an active space comprising 12 electrons distributed among 11
orbitals - i.e. CASSCF(12,11) - which provides a balanced description
of the multireference character along the reaction pathways.\\

\noindent
Starting from the existing 13877 geometries \cite{MM.h2coo:2025},
random displacements were applied to generate 1053 additional
structures in the three product regions (CO$_2$+H$_2$, H$_2$O+CO, and
HCO+OH), excluding configurations with energies more than 150 kcal/mol
above the optimized H$_2$COO. The resulting data set, combined with
the original structures, comprised a total of 14930 geometries.\\

\noindent
ML-PESs were trained using the PhysNet\cite{MM.physnet:2019} NN
architecture and an 80/10/10 split of the reference data set into
training, validation, and test subsets. Four independent NN-PESs were
trained, and the model with the smallest MAE$(E)$was selected for the
subsequent ML-MD simulations.\\

\begin{figure} [H]
    \centering
    \includegraphics[width=0.8\linewidth]{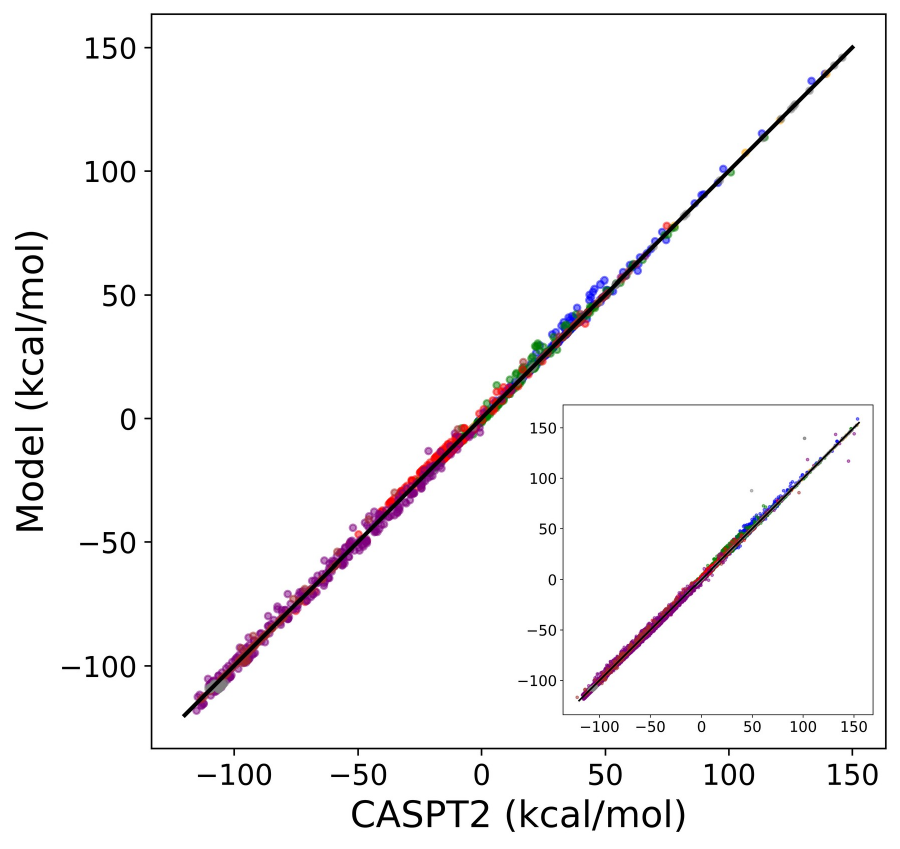}
    \caption{Relationship between the CASPT2/aVTZ reference energies
      and the model predictions. The main panel shows the test data
      set, comprising approximately 1500 structures. The species are
      color-coded as follows: blue, H$_2$COO; orange, HCO + OH; green,
      cyc-H$_2$CO$_2$; red, OCH$_2$O; purple, HCOOH; brown, CO$_2$ +
      H$_2$; and gray, H$_2$O + CO. The inset displays the training
      data set, which contains approximately 12000 structures. The
      coefficients of determination are $R^2 = 0.9984$ for the test
      set and $R^2 = 0.9983$ for the training set. The black solid
      line denotes the ideal 1:1 correlation.}
    \label{sifig:corr}
\end{figure}

\subsection{MD Simulations}
All production MD simulations were carried out using the pyCHARMM
software
package\cite{brooks2009charmm,buckner2023pycharmm,MM.charmm:2024} in
combination with interfaces to
PhysNet\cite{MM.pycharmm:2023,MM.asparagus:2025}. A time step of
$\Delta t = 0.1$ fs was employed throughout the simulations. The
initial conditions for all trajectories were prepared from a 1 ns
equilibrium $NVE$ simulation, following heating of H$_2$COO to 300 K
over 200 ps and subsequent equilibration for 50
ps.\cite{MM.h2coo:2024,MM.h2coo:2025}. From a set of 10 independent
simulations, coordinates and velocities were recorded every 100 fs,
yielding a total of $10^6$ structures to serve as starting points for
the non-equilibrium dynamics. Each trajectory was propagated up to a
maximum simulation time of 1 ns or they were terminated once
dissociation into molecular products had occurred. To assign
individual MD snapshots to specific states along the reaction pathway
(Figure \ref{fig:diagram}), a set of geometric criteria was applied,
summarized in Table \ref{sitab:criteria}.\\

\begin{table}[h!]
    \centering
    \begin{tabular}{c|c|c|c|c}
    \hline
    Species & H$_2$COO & cyc-H$_2$CO$_2$ & OCH$_2$O & HCOOH \\
    \hline
    distance CH & $<$ 1.8 & $<$ 1.5 & $<$ 1.5 & $<$ 1.5, $>$ 1.5 \\
    \hline
    distance CO & $<$ 1.8, $>$ 1.7 & $<$ 2.0 & $<$ 2.0 & $<$ 1.6, $<$ 1.8 \\
    \hline
    distance OO & $<$ 2.3 & n.u. & n.u. & $>$ 2.0 \\
    \hline
    angle COO & $>$ 90, $<$ 90 & $<$ 90 & $<$ 90 & $<$ 90 \\
    \hline
    angle OCO & $<$ 90 & $<$ 95 & $\geq$ 95 & $>$ 90 \\
    \hline
    \end{tabular}
    \caption{Geometrical criteria were used to assign structures along
      the trajectories. Distances and angles are reported in units of
      \AA\/ and $^\circ$, respectively. For the two C–H and C–O bond
      lengths, as well as the two possible C–O–O angles, the criteria
      were applied twice when necessary. The abbreviation “n.u.”
      denotes “not used.”}
    \label{sitab:criteria}
\end{table}

\noindent
The reactive dynamics was initiated by selectively exciting a
combination mode corresponding to 3 quanta in the CH-stretch
($3\nu_{\mathrm{CH}}$) and 1 quantum in the COO bending mode
($1\nu_{\mathrm{COO}}$), resulting in a total vibrational energy of
25.5 kcal/mol. This targeted vibrational excitation allows the system
to access the reactive regions of the PES efficiently and facilitates
barrier crossing toward product formation. Excitations closer to the
TS1-barrier energy lead to the same three dissociation products
(CO$_2$ + H$_2$, H$_2$O + CO, and HCO + OH), but the reactions occur
on significantly longer timescales.\cite{MM.h2coo:2025}\\

\noindent
Analysis of the fragment energies followed an established energy
partitioning scheme akin to that described earlier in the
literature.\cite{hase1998} For this, the total kinetic energy was
computed as the sum of atomic kinetic energies. Product fragments were
identified from the final atomic configuration, and for each fragment
the center-of-mass (CoM) velocity was obtained from mass-weighted
atomic velocities. The translational kinetic energy of each fragment
was defined as $E_{\rm trans} = \frac{1}{2} M V_{\mathrm{CoM}}^2$
where $M$ is the mass of the fragment, corresponding to three
translational degrees of freedom. After removing the CoM motion, the
internal kinetic energy $E_{\rm int} = E_{\rm tot} - E_{\rm trans}$
was evaluated using velocities relative to the fragment CoM.\\

\noindent
The instantaneous angular momentum was computed as $\mathbf{L} =
\sum_i m_i \left( \mathbf{r}_i - \mathbf{r}_{\mathrm{CoM}} \right)
\times \left( \mathbf{v}_i - \mathbf{v}_{\mathrm{CoM}} \right)$ and
the moment-of-inertia tensor as $\mathbf{I} = \sum_i m_i \left[ \left|
  \mathbf{r}_i - \mathbf{r}_{\mathrm{COM}} \right|^2 \mathbf{1} -
  \left( \mathbf{r}_i - \mathbf{r}_{\mathrm{CoM}} \right) \left(
  \mathbf{r}_i - \mathbf{r}_{\mathrm{CoM}} \right)^{\mathrm{T}}
  \right]$. From this, the angular velocity ${\bf \omega}$ was
obtained by solving $\mathbf{I} \boldsymbol{\omega} = \mathbf{L}$ and
the rotational kinetic energy was computed as $E_{\mathrm{rot}} =
\frac{1}{2} \sum_i m_i \left| \boldsymbol{\omega} \times \left(
\mathbf{r}_i - \mathbf{r}_{\mathrm{COM}} \right) \right|^2$
corresponding to two rotational degrees of freedom for linear
fragments and three for nonlinear fragments. Finally, the vibrational
kinetic energy was obtained from $E_{\mathrm{vib}} = E_{\mathrm{int}}
- E_{\mathrm{rot}}$ where the internal kinetic energy is given
by$E_{\mathrm{int}} = \frac{1}{2} \sum_i m_i \left| \mathbf{v}_i -
\mathbf{v}_{\mathrm{CoM}} \right|^2$ The vibrational term accounts for
the remaining $3N - 5$ (linear fragments) or $3N - 6$ (nonlinear
fragments) degrees of freedom.\\

\subsection{Product State PESs}
\begin{figure}[h!]
    \centering \includegraphics[width=0.8\linewidth]{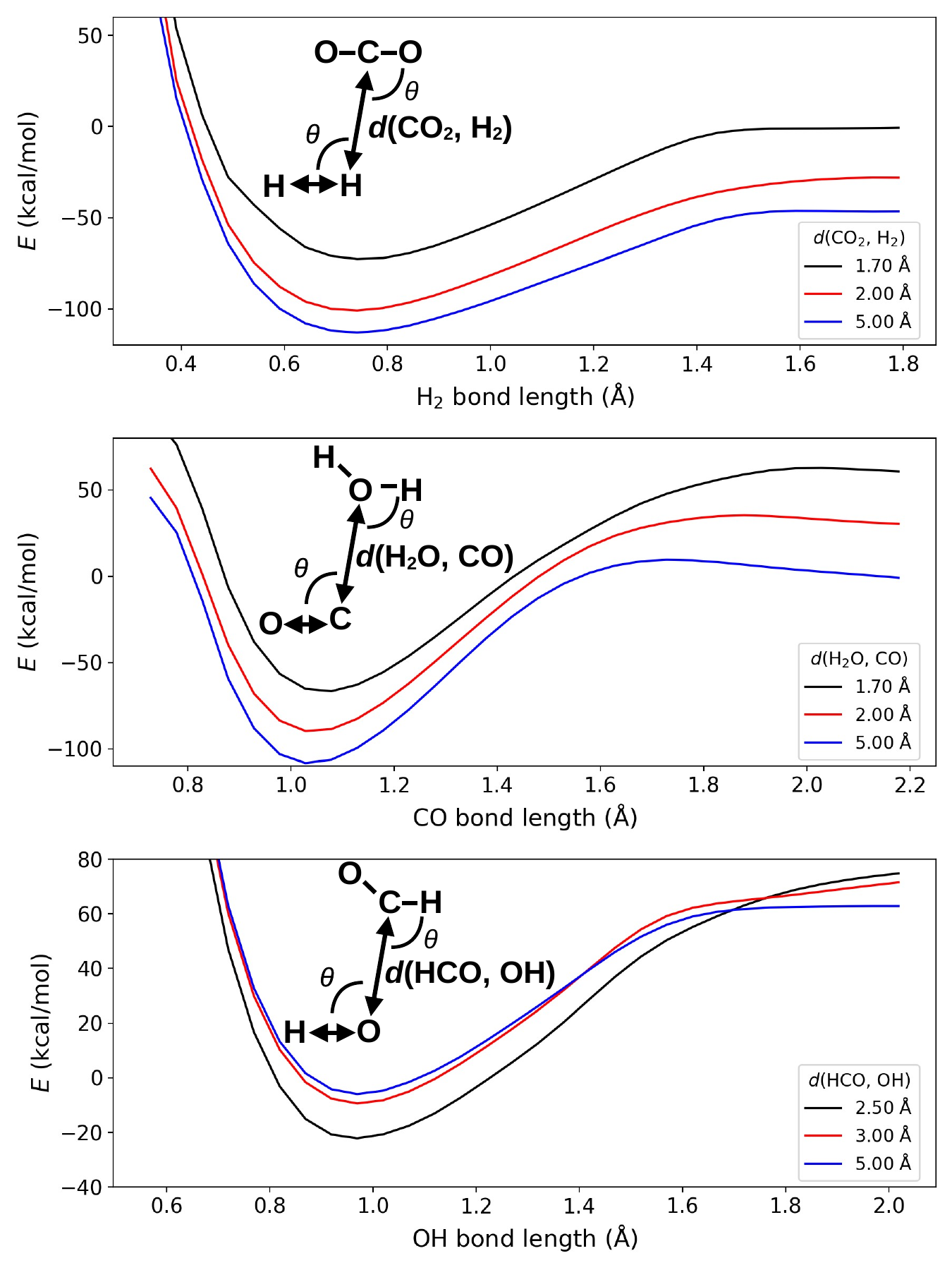}
    \caption{The energy profile of the diatomic products, i.e., H$_2$
      (upper panel), CO (middle panel), and OH (lower panel). For each
      species, the diatom and the corresponding triatomic molecule
      were placed at three different distances, as indicated by the
      different colors in this figure. All the values of angle
      $\theta$ are fixed at $100^{\circ}$. The structures of the
      triatomic molecules were fixed at their equilibrium structure,
      while the bond length in the diatomic was scanned around its
      equilibrium value.}
    \label{sifig:diatom}
\end{figure}

For H$_2$, CO, and OH, the energy profiles as functions of bond length
are only modestly perturbed as the distance to the triatomic partner
changes, see Figure \ref{sifig:diatom}. This indicates that the
intramolecular bond dynamics of the diatomic products are largely
decoupled from the translational separation of the fragments,
especially at larger separations. The angle between the diatomic and
the triatomic is fixed at $100^{\circ}$. At large fragment separations
($\sim 5$ \AA, corresponding to the blue curves in Figure
\ref{sifig:diatom}), the interaction effectively terminates, and the
energy profiles approach those of the isolated diatomic
species. Different colors represent different distances between the
diatomic and triatomic fragments, highlighting how the interaction
energy changes with separation. In the asymptotic limit, a crossing
appears for the HCO + OH channel. However, this feature occurs at
energies well above the initial energy of the simulations
(approximately 25.5 kcal/mol) and therefore does not influence the MD
trajectories, considering that the initial energy of simulation is
around 25.5 kcal/mol.\\

\begin{figure}[h!]
    \centering \includegraphics[width=0.8\linewidth]{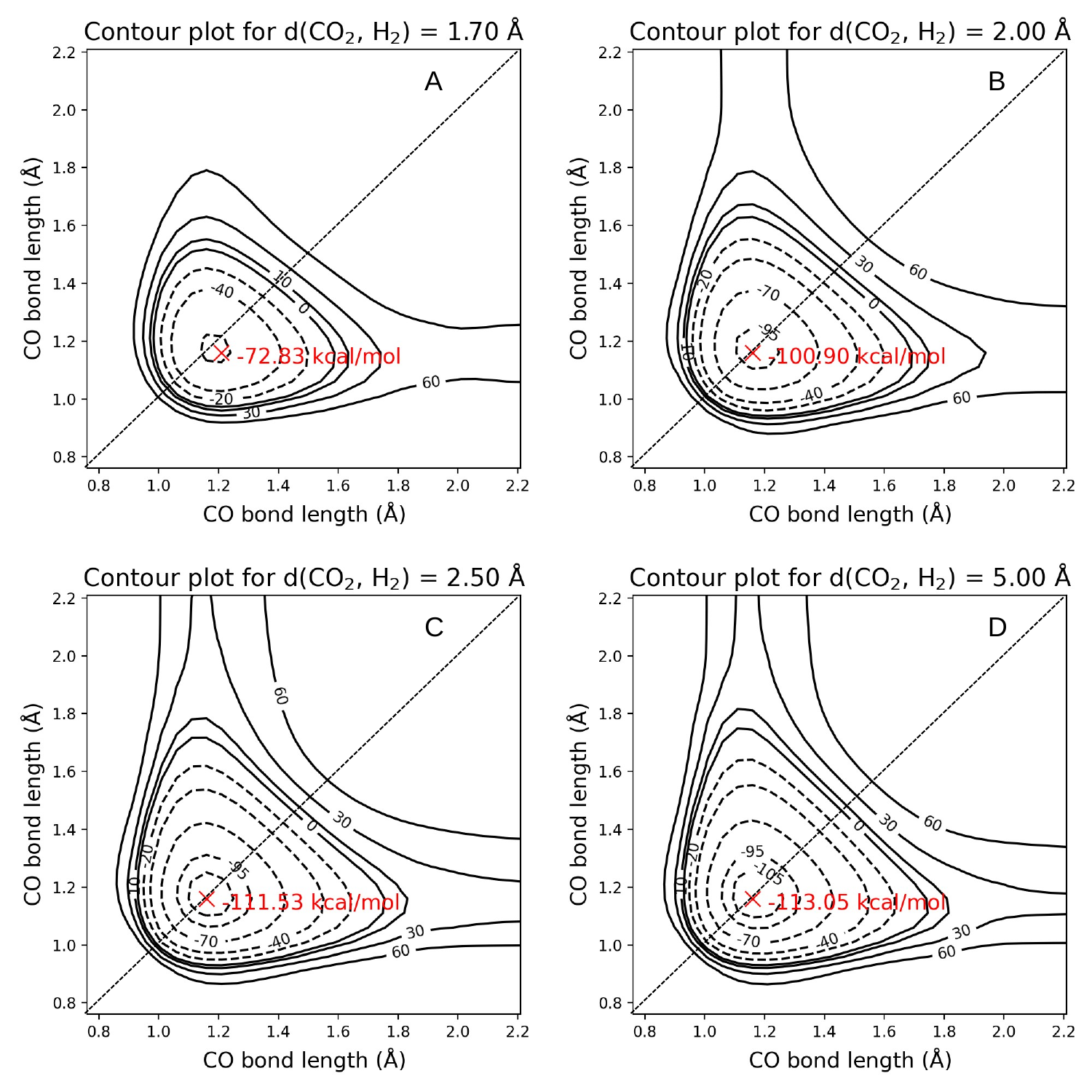}
    \caption{The PES $V(r_{\rm CO_A},r_{\rm CO_B})$ of the CO$_2$
      product with the partner diatomic fragment H$_2$ was fixed at
      its equilibrium bond length of 0.741 \AA\/ whereas the CO$_2$
      molecule was constrained to remain linear. In each panel, a
      specific CO$_2$–H$_2$ separation is examined. The definition of
      this distance and the position of H$_2$ relative to CO$_2$ are
      presented in the upper panel of Figure \ref{sifig:diatom}.}
    \label{sifig:cont_co2}
\end{figure}

\begin{figure}[H]
    \centering \includegraphics[width=0.8\linewidth]{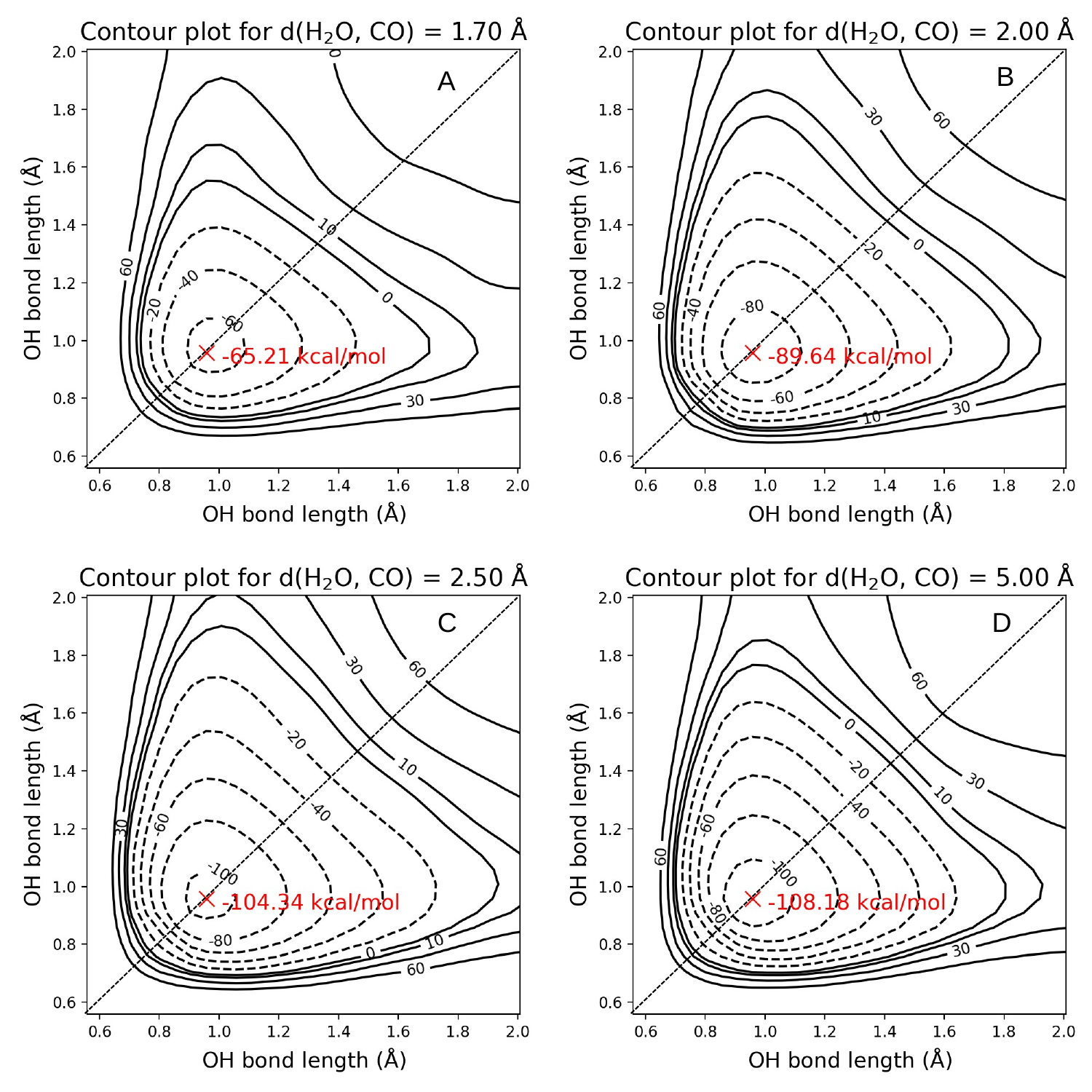}
    \caption{The PES $V(r_{\rm OH_A},r_{\rm OH_B})$ of the triatomic
      product H$_2$O was evaluated while the partner diatomic fragment
      CO was fixed at its equilibrium bond length of 1.128 \AA\/. The
      HOH angle in H$_2$O molecule was constrained to 104.5$^{\circ}$.
      In each panel, a certain H$_2$O–CO separation distance is
      examined. The precise definition of this distance and the
      position of CO relative to H$_2$O are presented in the middle
      panel of Figure \ref{sifig:diatom}. The two OH bond lengths in
      H$_2$O are varied around their equilibrium value.}
    \label{sifig:cont_h2o}
\end{figure}

\begin{figure}[H]
    \centering
    \includegraphics[width=0.8\linewidth]{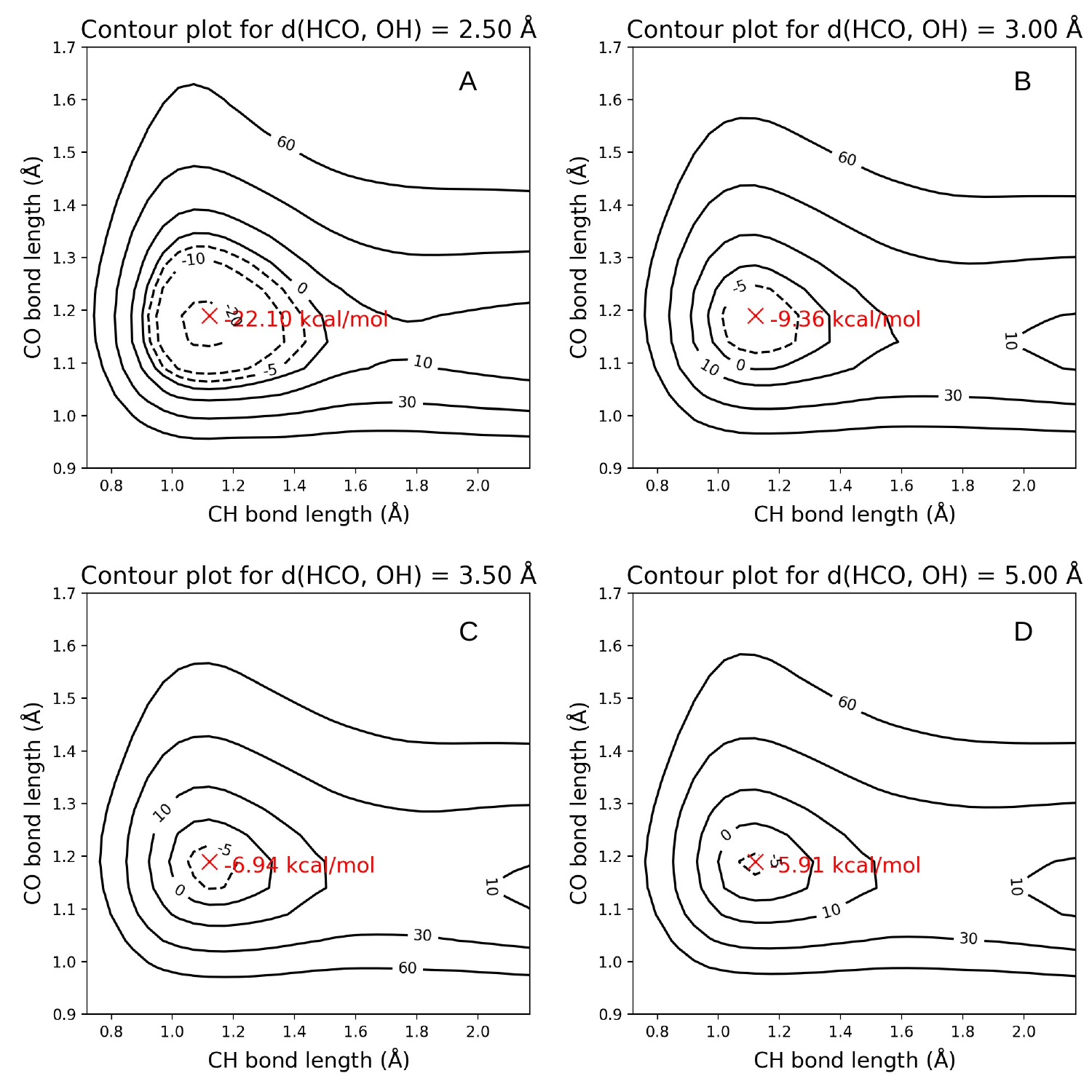}
    \caption{The PES $V(r_{\rm CH},r_{\rm CO})$ of the triatomic
      product HCO with the diatomic fragment OH fixed at its
      equilibrium bond length of 0.969 \AA\/. The H-C-O angle for HCO
      was constrained to 124$^{\circ}$. In each panel, a HCO–OH
      separations ranging from 2.5 \AA\/ to 5.0 \AA\/ are examined,
      see the lower panel of Figure \ref{sifig:diatom}.}
    \label{sifig:cont_hco}
\end{figure}

\noindent
Figures \ref{sifig:cont_co2} to \ref{sifig:cont_hco} illustrate the PESs
of the triatomic products CO$_2$, H$_2$O, and HCO in the presence of
their corresponding diatomic partners, H$_2$, CO, and OH,
respectively. The two CO bond lengths in CO$_2$ are varied around
their equilibrium value. At short diatomic-triatomic distance, as the
upper left panel shows, the energy profile of two CO bond stretching
is not symmetric, since the interaction between CO$_2$ and H$_2$ is
not symmetric, see Figure \ref{sifig:diatom}. The close agreement
between the C--O dissociation energy shown in this figure
(approximately 130 kcal/mol) and the experimentally reported value of
126 kcal/mol\cite{gong2024bond} supports the reliability of the
present PES. For CO$_2$ + H$_2$, the CO bonds remain intact because
dissociation requires much higher energy and in the MD trajectories
some energy is carried away by H$_2$, making CO bond cleavage
unlikely.\\

\noindent
For the H$_2$O + CO system, variations in the OH bond lengths are
similarly explored across different H$_2$O--CO separations, revealing
a stable triatomic product region under the simulation conditions.\\

\noindent
For HCO + OH, the potential energy minimum rises with increasing
HCO–OH separation, reflecting the barrierless nature of HCOOH
decomposition. The CH bond dissociation energy of approximately 15
kcal/mol is below the total initial energy of the simulation, allowing
a subset of trajectories to access the H + CO + OH channel. The CH and
CO bond lengths in HCO are varied around their equilibrium value. At a
separation of 5 \AA\/, the HCO molecule should behave essentially as a
free HCO radical. The CH bond dissociation energy is $\sim 15$
kcal/mol, in good agreement with previous studies reporting values of
13.36 kcal/mol\cite{peters:2013} and 14.40
kcal/mol\cite{vichietti:2020}. In addition, along the H-atom
dissociation pathway, a transition state is observed in this
figure. This transition state was also identified in earlier
MRCI/aug-cc-pVQZ\cite{peters:2013} and
CCSD(T)/CBS//CCSD/aug-cc-pVTZ\cite{vichietti:2020} studies and lies
below the initial 25.5 kcal/mol of energy available in the MD
simulations. As a result, a small fraction of trajectories proceeded
to form the H + CO + OH products. The minimum rises with increasing
$d{\rm (HCO, OH)}$, in contrast to the CO$_2$+H$_2$ and H$_2$O+CO
cases. This behavior is expected, as HCOOH decomposition proceeds
without a transition state. The potential energy rises as the HCO and
OH separation grows. These energy profiles collectively demonstrate
how the interaction between triatomic products and their diatomic
partners shapes the accessible reaction pathways in the MD
simulations.\\

\begin{figure} [H]
    \centering \includegraphics[width=0.8\linewidth]{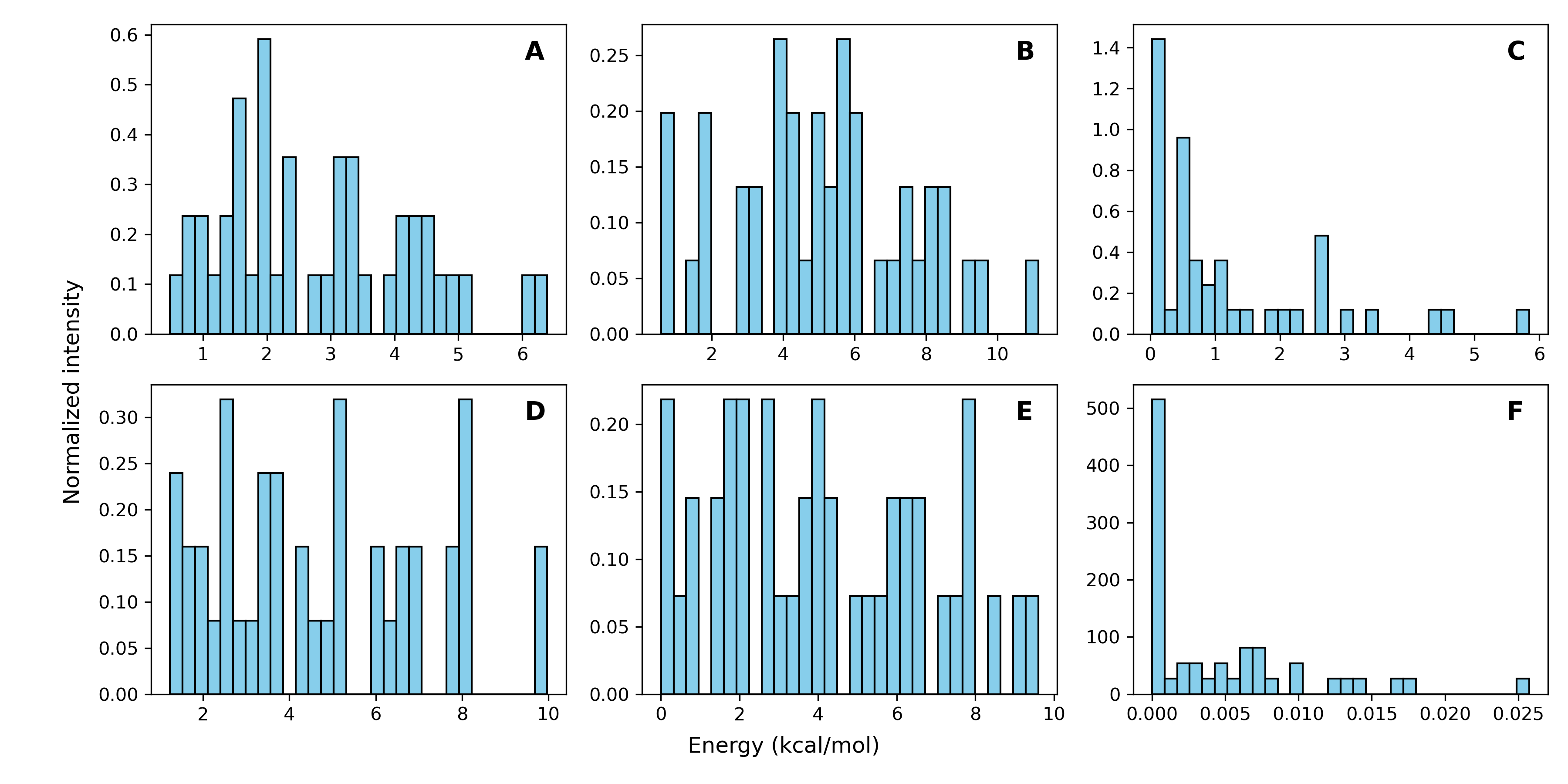}
    \caption{Energy distributions of the product HCO+OH from 40
      simulations. The top panels show fragment HCO and bottom show
      fragment OH. From left to right, the distributions correspond to
      translational, translational, rotational, and vibrational
      energies. Because only 40 trajectories produce this product, the
      data are not sufficient to yield a converged energy
      distribution, as was achieved for CO$_2$+H$_2$ and H$_2$O+CO.}
    \label{sifig:hco_oh_his}
\end{figure}

\begin{figure} [H]
    \centering \includegraphics[width=0.8\linewidth]{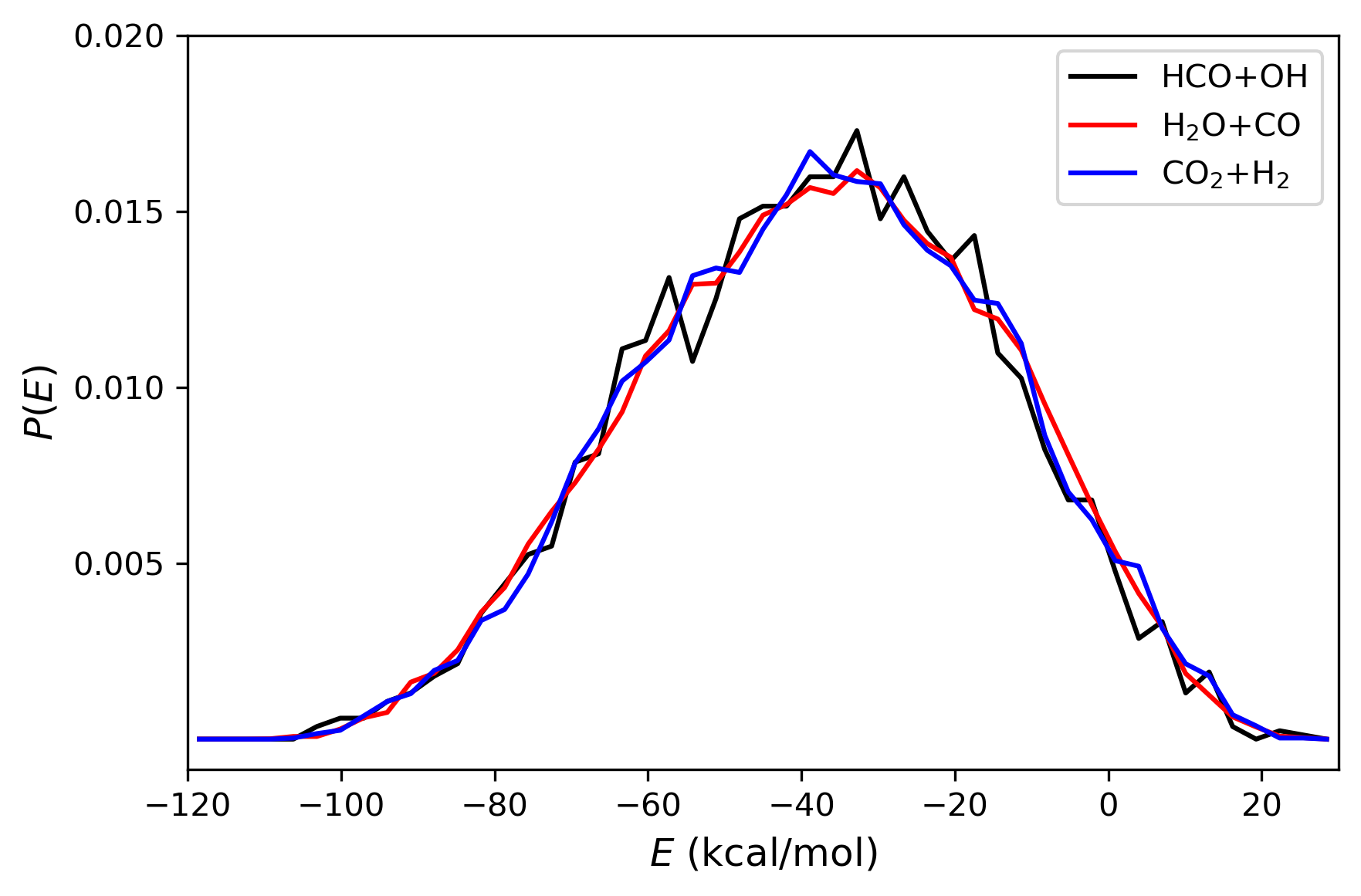}
    \caption{The energy distributions of HCOOH associated with the
      three product channels, HCO+OH (black), H$_2$O+CO (red), and
      CO$_2$+H$_2$ (blue). The peak is located at $\sim -40$ kcal/mol,
      whereas the potential energy of equilibrium HCOOH is around
      --140 kcal/mol. This indicates that during the simulation HCOOH
      possesses $\sim 100$ kcal/mol of internal energy.}
    \label{sifig:fa_e}
\end{figure}

\begin{figure}[h!]
    \centering
    \includegraphics[width=1.0\linewidth]{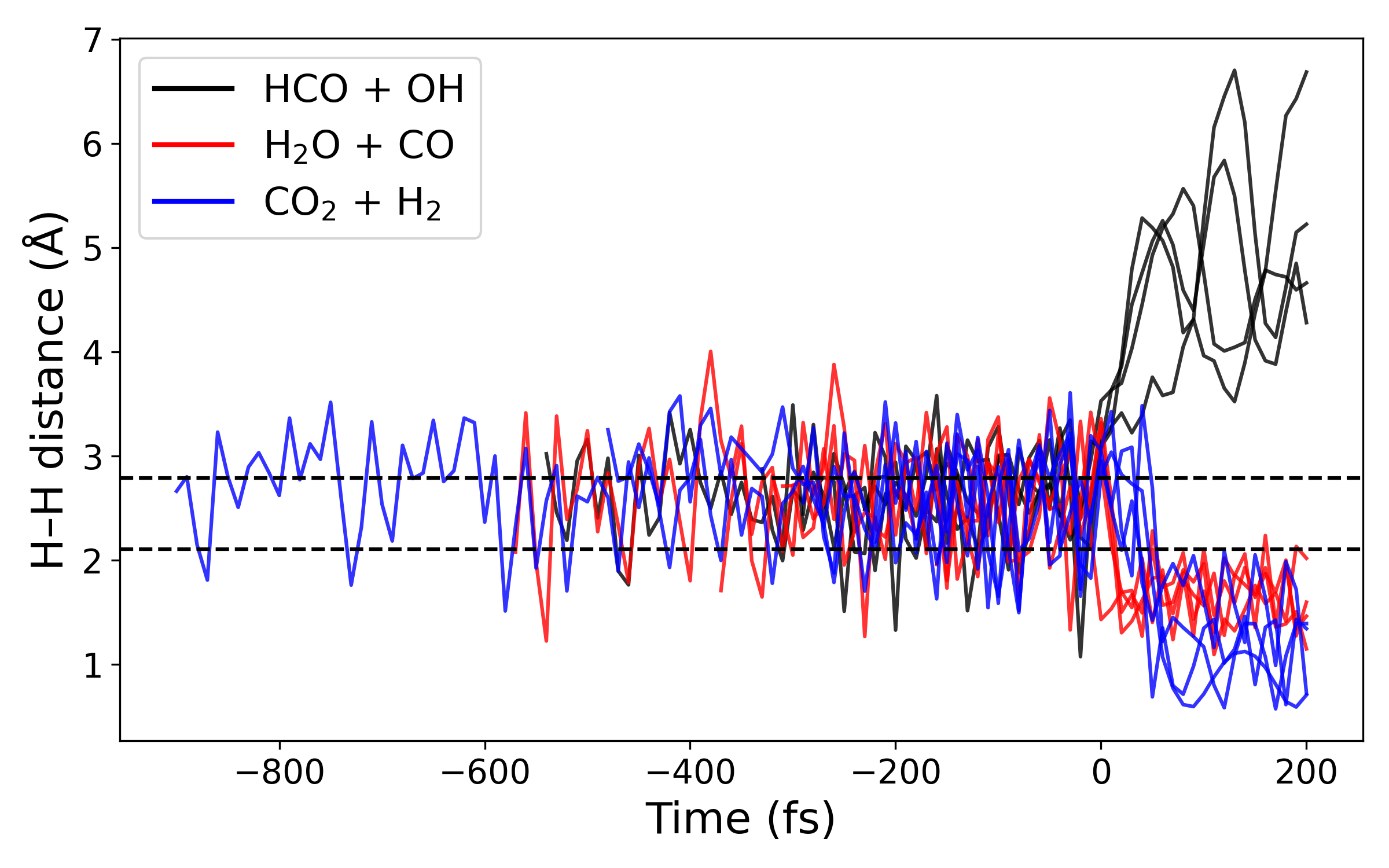}
    \caption{Time evolution of the H–H separation along reactive
      molecular dynamics trajectories leading to three different
      product channels. Trajectories forming HCO + OH are shown in
      black, H$_2$O + CO in red, and CO$_2$ + H$_2$ in blue, with four
      representative trajectories shown for each channel. The time
      axis is shifted such that $t=0$ corresponds to the frame of
      product formation; negative times indicate configurations in the
      HCOOH region, while positive times correspond to the product
      region. The two dashed lines indicate the H–H separation in the
      equilibrium structures of {\it cis-} and {\it trans-} HCOOH,
      respectively.}
    \label{sifig:HH_distances}
\end{figure}

\clearpage
\bibliography{refs.clean}

@string{jcp = "J. Chem. Phys."}

@string{cpl = "Chem. Phys. Lett."}

@string{cphysc = "Comp. Phys. Comm."}

@string{csr = "Chem. Soc. Rev."}

@string{arpc = "Ann. Rev. Phys. Chem."}

@string{jpca = "J. Phys. Chem. A"}

@string{jpcb = "J. Phys. Chem. B"}

@string{jctc = "J. Chem. Theory. Comput."}

@string{prl = "Phys. Rev. Lett."}

@string{jcc = "J. Comput. Chem."}

@string{jpcl = "J. Phys. Chem. Lett."}

@string{jlac= "Justus Liebigs Ann. Chem."}

@string{jcc = "J. Chem. Comp."}

@string{ic = "Inorg. Chem."}

@string{anie = "Angew. Chem. Int. Ed."}

@string{pnas = "Proc. Nat. Acad. Sci. USA"}

@string{bc = "Biochemistry"}

@string{proteins = "Proteins"}

@string{nc = "Nat. Chem."}

@string{sci = "Science"}

@string{nature = "Nature"}

@string{pccp = "Phys. Chem. Chem. Phys."}

@string{BC = {Biochemistry}}

@string{JCC = {J. Comput. Chem.}}

@string{JCP = {J. Chem. Phys.}}

@string{JCTC = {J. Chem. Theory Comput.}}

@string{JPCB = {J. Phys. Chem. B}}

@string{M = {Microbiology}}

@string{PNAS = {Proc. Natl. Acad. Sci. USA}}

@string{PRL = {Phys. Rev. Lett.}}

@string{S = {Science}}

@article{MM.physnet:2019,
	title        = {PhysNet: A Neural Network for Predicting
                  Energies, Forces, Dipole Moments, and Partial
                  Charges},
	author       = {Unke, Oliver T. and Meuwly, Markus},
	year         = 2019,
	journal      = jctc,
	volume       = 15,
	number       = 6,
	pages        = {3678--3693}
}

@article{criegee1949ozonisierung,
  title={Die Ozonisierung des 9, 10-Oktalins},
  author={Criegee, Rudolf and Wenner, Gotthilf},
  journal=jlac,
  volume={564},
  number={1},
  pages={9--15},
  year={1949},
  publisher={Wiley Online Library}
}

@article{yin2017does,
  title={How does substitution affect the unimolecular reaction rates of Criegee intermediates?},
  author={Yin, Cangtao and Takahashi, Kaito},
  journal=pccp,
  volume={19},
  number={19},
  pages={12075--12084},
  year={2017},
  publisher={Royal Society of Chemistry}
}

@article{vereecken2017unimolecular,
  title={Unimolecular decay strongly limits the atmospheric impact of Criegee intermediates},
  author={Vereecken, L and Novelli, A and Taraborrelli, D},
  journal=pccp,
  volume={19},
  number={47},
  pages={31599--31612},
  year={2017},
  publisher={Royal Society of Chemistry}
}

@article{green2017selective,
  title={Selective deuteration illuminates the importance of tunneling in the unimolecular decay of Criegee intermediates to hydroxyl radical products},
  author={Green, Amy M and Barber, Victoria P and Fang, Yi and Klippenstein, Stephen J and Lester, Marsha I},
  journal=pnas,
  volume={114},
  number={47},
  pages={12372--12377},
  year={2017},
  publisher={National Acad Sciences}
}

@article{brooks2009charmm,
  title={CHARMM: the biomolecular simulation program},
  author={Brooks, Bernard R and Brooks III, Charles L and Mackerell Jr, Alexander D and Nilsson, Lennart and Petrella, Robert J and Roux, Beno{\^\i}t and Won, Youngdo and Archontis, Georgios and Bartels, Christian and Boresch, Stefan and others},
  journal=jcc,
  volume={30},
  number={10},
  pages={1545--1614},
  year={2009},
  publisher={Wiley Online Library}
}

@article{lester:2016,
  title={Unimolecular dissociation dynamics of vibrationally activated
                  CH$_3$CHOO Criegee intermediates to OH radical
                  products},
  author={Kidwell, Nathanael M and Li, Hongwei and Wang, Xiaohong and
                  Bowman, Joel M and Lester, Marsha I},
  journal=nc,
  volume={8},
  number={5},
  pages={509--514},
  year={2016},
  publisher={Nature Publishing Group}
}

@article{fang:2016,
Author = {Fang, Yi and Liu, Fang and Barber, Victoria P. and
                  Klippenstein, Stephen J. and McCoy, Anne B. and
                  Lester, Marsha I.},
Title = {{Communication: Real time observation of unimolecular decay
                  of Criegee intermediates to OH radical products}},
Journal = jcp,
Year = {{2016}},
Volume = {{144}},
pages={061102},
Number = {{6}},
Article-Number = {{061102}},
}

@article{MM.mbno:2016,
  title={Structural Interpretation of Metastable States in Myoglobin--NO},
  author={Soloviov, Maksym and Das, Akshaya K and Meuwly, Markus},
  journal=anie,
  volume={55},
  number={34},
  pages={10126--10130},
  year={2016},
  publisher={Wiley Online Library}
}

@article{molpro:2020,
Author = {Werner, Hans-Joachim and Knowles, Peter J. and Manby,
                  Frederick R. and Black, Joshua A. and Doll, Klaus
                  and Hesselmann, Andreas and Kats, Daniel and Koehn,
                  Andreas and Korona, Tatiana and Kreplin, David
                  A. and Ma, Qianli and Miller, III, Thomas F. and
                  Mitrushchenkov, Alexander and Peterson, Kirk A. and
                  Polyak, Iakov and Rauhut, Guntram and Sibaev, Marat},
Title = {{The Molpro quantum chemistry package}},
Journal = jcp,
Year = {{2020}},
Volume = {{152}},
Number = {{14}},
Pages = {{144107}},
}

@article{fang:2016deep,
  title={Deep tunneling in the unimolecular decay of CH$_3$CHOO
                  Criegee intermediates to OH radical products},
  author={Fang, Yi and Liu, Fang and Barber, Victoria P and
                  Klippenstein, Stephen J and McCoy, Anne B and
                  Lester, Marsha I},
  journal=JCP,
  volume={145},
  number={23},
  pages={234308},
  year={2016},
  publisher={AIP Publishing LLC}
}

@article{MM.criegee:2021,
  title={Thermal and Vibrationally Activated Decomposition of the
                  syn-CH$_3$CHOO Criegee Intermediate},
  author={Upadhyay, Meenu and Meuwly, Markus},
  journal={ACS Earth Space Chem.},
  volume={5},
  number={12},
  pages={3396--3406},
  year={2021},
  publisher={ACS Publications}
}

@article{MM.pycharmm:2023,
  title={PhysNet Meets CHARMM: A Framework for Routine Machine
                  Learning/Molecular Mechanics Simulations},
  author={Song, Kaisheng and K{\"a}ser, Silvan and T{\"o}pfer, Kai and
                  Vazquez-Salazar, Luis Itza and Meuwly, Markus},
  journal=jcp,
volume={159},
pages={024125},
  year={2023}
}

@article{austin:1975,
  title={Dynamics of ligand binding to myoglobin},
  author={Austin, R and Beeson, KW and Eisenstein, L and Frauenfelder,
                  H and Gunsalus, IC},
  journal=bc,
  volume={14},
  number={24},
  pages={5355--5373},
  year={1975},
  publisher={ACS Publications}
}

@article{welz:2012,
  title={Direct kinetic measurements of Criegee intermediate (CH$_2$OO)
                  formed by reaction of CH$_2$I with O$_2$},
  author={Welz, Oliver and Savee, John D and Osborn, David L and Vasu,
                  Subith S and Percival, Carl J and Shallcross, Dudley
                  E and Taatjes, Craig A},
  journal=S,
  volume={335},
  number={6065},
  pages={204--207},
  year={2012},
  publisher={American Association for the Advancement of Science}
}

@article{lester:2024,
  title={Nonstatistical Unimolecular Decay of the CH{$_2$}OO Criegee
                  Intermediate in the Tunneling Regime},
  author={Qian, Yujie and Nguyen, Thanh Lam and Franke, Peter R and
                  Stanton, John F and Lester, Marsha I},
  journal=jpcl,
  volume={15},
  pages={6222--6229},
  year={2024},
  publisher={ACS Publications}
}

@article{MM.h2coo:2024,
  title={OH-Formation following vibrationally induced reaction
                  dynamics of H$_2$COO},
  author={Song, Kaisheng and Upadhyay, Meenu and Meuwly, Markus},
  journal=pccp,
  volume={26},
  number={16},
  pages={12698--12708},
  year={2024},
  publisher={Royal Society of Chemistry}
}

@article{buckner2023pycharmm,
  title={pyCHARMM: embedding CHARMM functionality in a python framework},
  author={Buckner, Joshua and Liu, Xiaorong and Chakravorty, Arghya and Wu, Yujin and Cervantes, Luis F and Lai, Thanh T and Brooks III, Charles L},
  journal={J. Chem. Theory Comput.},
  volume={19},
  number={12},
  pages={3752--3762},
  year={2023},
  publisher={ACS Publications}
}

@article{stanton:2015,
  title={Stabilization of the simplest Criegee intermediate from the
                  reaction between ozone and ethylene: A high-level
                  quantum chemical and kinetic analysis of ozonolysis},
  author={Nguyen, Thanh Lam and Lee, Hyunwoo and Matthews, Devin A and
                  McCarthy, Michael C and Stanton, John F},
  journal=jpca,
  volume={119},
  number={22},
  pages={5524--5533},
  year={2015},
  publisher={ACS Publications}
}

@article{MM.asparagus:2025,
  author    = {Kilian Töpfer and Luis I. Vázquez-Salazar and Markus
                  Meuwly},
  title     = {{Asparagus: A Toolkit for Autonomous, User-Guided
                  Construction of Machine-Learned Potential Energy
                  Surfaces}},
  journal   = cphysc,
  year      = {2025},
  volume    = {308},
  pages     = {109446},
  publisher={Elsevier}
}

@article{MM.charmm:2024,
  author    = {Wonmuk Hwang and Steven L. Austin and Arnaud Blondel
                  and Eric D. Boittier and Stefan Boresch and Matthias
                  Buck and Joshua Buckner and Amedeo Caflisch and
                  Hao‑Ting Chang and Xi Cheng and Yeol Kyo Choi and
                  Jhih‑Wei Chu and Michael F. Crowley and Qiang Cui
                  and Ana Damjanovic and Yuqing Deng and Mike Devereux
                  and Xinqiang Ding and Michael F. Feig and Jiali Gao
                  and David R. Glowacki and James E. Gonzales II and
                  Mehdi B. Hamaneh and Edward D. Harder and Ryan
                  L. Hayes and Jing Huang and Yandong Huang and
                  Phillip S. Hudson and Wonpil Im and Shahidul
                  M. Islam and Wei Jiang and Michael R. Jones and
                  Silvan Käser and Fiona L. Kearns and Nathan R. Kern
                  and Jeffery B. Klauda and Themis Lazaridis and
                  Jinhyuk Lee and Justin A. Lemkul and Xiaorong Liu
                  and Yun Luo and Alexander D. MacKerell Jr. and Dan
                  T. Major and Markus Meuwly and Kwangho Nam and
                  Lennart Nilsson and Victor Ovchinnikov and Emanuele
                  Paci and Soohyung Park and Richard W. Pastor and
                  Amanda R. Pittman and Carol Beth Post and Sam Prasad
                  and Jingzhi Pu and Yifei Qi and Thenmalarchelvi
                  Rathinavelan and Daniel R. Roe and Benoit Roux and
                  Christopher N. Rowley and Jana Shen and Andrew
                  C. Simmonett and Alexander J. Sodt and Kai Töpfer
                  and Meenu Upadhyay and Arjan van der Vaart and Luis
                  I. Vázquez‑Salazar and Richard M. Venable and Luke
                  C. Warrensford and H. Lee Woodcock and Yujin Wu and
                  Charles L. Brooks III and Bernard R. Brooks and
                  Martin Karplus},
  title     = {{CHARMM at 45: Enhancements in Accessibility, Functionality, and Speed}},
  journal   = jpcb,
  year      = {2024},
  volume    = {128},
  number    = {41},
  pages     = {9976–10042},
  doi       = {10.1021/acs.jpcb.4c04100},
}

@article{MM.criegee:2023,
  title={Molecular Simulation for Atmospheric Reactions:
                  Non-Equilibrium Dynamics, Roaming, and
                  Glycolaldehyde Formation following Photoinduced
                  Decomposition of syn-Acetaldehyde Oxide},
  author={Upadhyay, Meenu and To\"opfer, Kai and Meuwly, Markus},
  journal=jpcl,
  volume={15},
  pages={90--96},
  publisher={ACS Publications}
}

@article{peters:2013,
  title={The H+ CO $\leftrightarrow$ HCO reaction studied by ab initio
                  benchmark calculations},
  author={Peters, Phillip S and Duflot, Denis and Wiesenfeld, Laurent
                  and Toubin, C{\'e}line},
  journal=jcp,
  volume={139},
  number={16},
  year={2013},
  publisher={AIP Publishing}
}

@article{vichietti:2020,
  title={Accurate rate constants for the forward and reverse H+
                  CO$\leftrightarrow$HCO reactions at the
                  high-pressure limit},
  author={Vichietti, Rafael M and Machado, Francisco BC and Haiduke, Roberto LA},
  journal={ACS omega},
  volume={5},
  number={37},
  pages={23975--23982},
  year={2020},
  publisher={ACS Publications}
}

@article{stone2014kinetics,
  title={Kinetics of CH$_2$OO reactions with SO$_2$, NO$_2$, NO,
                  H$_2$O and CH$_3$CHO as a function of pressure},
  author={Stone, Daniel and Blitz, Mark and Daubney, Laura and Howes,
                  Neil UM and Seakins, Paul},
  journal=pccp,
  volume={16},
  number={3},
  pages={1139--1149},
  year={2014},
  publisher={Royal Society of Chemistry}
}

@article{qiu2019detection,
  title={Detection of the simplest Criegee intermediate CH$_2$OO in
                  the $\nu$4 band using a continuous wave quantum
                  cascade laser and its kinetics with SO$_2$ and
                  NO$_2$},
  author={Qiu, Junting and Tonokura, Kenichi},
  journal=cpl,
  volume={737},
  pages={100019},
  year={2019},
  publisher={Elsevier}
}

@article{yin2024revealing,
  title={Revealing new pathways for the reaction of Criegee
                  intermediate CH$_2$OO with SO$_2$},
  author={Yin, Cangtao and Czak{\'o}, G{\'a}bor},
  journal={Comm. Chem.},
  volume={7},
  number={1},
  pages={157},
  year={2024},
  publisher={Nature Publishing Group UK London}
}

@article{MM.h2coo:2025,
  title = {Photodissociation dynamics of energized H$_2$COO: Formation
                  of molecular products},
  author = {Yin, Cangtao and K{\"a}ser, Silvan and Upadhyay, Meenu and Meuwly, Markus},
  journal = jcp,
  volume = {163},
  number = {21},
  pages = {214305},
  year = {2025},
}

@article{yin2018effect,
  title={Effect of unsaturated substituents in the reaction of Criegee
                  intermediates with water vapor},
  author={Yin, Cangtao and Takahashi, Kaito},
  journal=pccp,
  volume={20},
  number={30},
  pages={20217--20227},
  year={2018},
  publisher={Royal Society of Chemistry}
}

@article{chao2015direct,
  title={Direct kinetic measurement of the reaction of the simplest
                  Criegee intermediate with water vapor},
  author={Chao, Wen and Hsieh, Jun-Ting and Chang, Chun-Hung and Lin, Jim Jr-Min},
  journal=sci,
  volume={347},
  number={6223},
  pages={751--754},
  year={2015},
  publisher={American Association for the Advancement of Science}
}

@article{gong2024bond,
  title={Bond dissociation energy of CO2 with spectroscopic accuracy using state-to-state resolved threshold fragment yield spectra},
  author={Gong, Shiyan and Wang, Peng and Mo, Yuxiang},
  journal=jpcl,
  volume={15},
  number={43},
  pages={10842--10848},
  year={2024},
  publisher={ACS Publications}
}

@book{hase1998,
	title        = {Encyclopedia of Computational Chemistry},
	author       = {Hase, W. L.},
	year         = 1998,
	publisher    = {Wiley, New York},
    pages        = {399--407}
}

@article{kohlrausch:1854,
  author  = {Kohlrausch, Rudolf},
  title   = {Theorie des elektrischen R{\"u}ckstandes in der Leidener Flasche},
  journal = {Annalen der Physik und Chemie},
  year    = {1854},
  volume  = {91},
  pages   = {179--214}
}

@article{watts:1970,
  author  = {Williams, G. and Watts, D. C.},
  title   = {Non-symmetrical dielectric relaxation behaviour arising from a simple empirical decay function},
  journal = {Transactions of the Faraday Society},
  year    = {1970},
  volume  = {66},
  pages   = {80--85},
  doi     = {10.1039/TF9706600080}
}

@article{austin:1974,
  author  = {Austin, R. H. and Beeson, K. W. and Eisenstein, L. and
                  Frauenfelder, H. and Gunsalus, I. C.},
  title   = {Dynamics of Ligand Binding to Myoglobin},
  journal = prl,
  year    = {1974},
  volume  = {32},
  number  = {8},
  pages   = {403--405},
}

@article{richert:2002,
  author  = {Richert, Ranko},
  title   = {Heterogeneous dynamics in liquids: fluctuations in space and time},
  journal = {Journal of Physics: Condensed Matter},
  year    = {2002},
  volume  = {14},
  number  = {23},
  pages   = {R703--R738},
  doi     = {10.1088/0953-8984/14/23/201}
}

@article{bauchy:2015,
  author  = {Yu, Yingtian and Wang, Mengyi and Zhang, Dawei and
                  Bauchy, Mathieu},
  title   = {Stretched exponential relaxation of glasses at low
                  temperature},
  journal = prl,
  year    = {2015},
  volume  = {115},
  pages   = {165901},
}

@article{MM.o2:2019,
  author    = {Marco Pezzella and Markus Meuwly},
  title     = {O$_2$ formation in cold environments},
  journal   = pccp,
  year      = {2019},
  volume    = {21},
  number    = {11},
  pages     = {5559--5571},
  doi       = {10.1039/C8CP07474G},
}

@article{wolynes:1991,
  author  = {Frauenfelder, Hans and Sligar, Stephen G. and Wolynes,
                  Peter G.},
  title   = {The energy landscapes and motions of proteins},
  journal = sci,
  year    = {1991},
  volume  = {254},
  number  = {5038},
  pages   = {1598--1603},
}

@article{sun:2024,
  author  = {Sun, Ge and Qin, Yuan and Zheng, Xianfeng and Song, Yu
                  and Zhang, Jingsong},
  title   = {Photodissociation of HCO Radical via the
                  $\tilde{A}^{2}A''$ State: Accurate Determination of
                  Bond Dissociation Energy $D_{0}$(H--CO)},
  journal = {Chin. J. Chem. Phys.},
  volume  = {37},
  number  = {6},
  pages   = {857--862},
  year    = {2024},
  doi     = {10.1063/1674-0068/cjcp2411146}
}

@article{francisco:2012,
  author  = {Vereecken, Luc and Francisco, Joseph S.},
  title   = {Theoretical studies of atmospheric reaction mechanisms in
                  the troposphere},
  journal = csr,
  year    = {2012},
  volume  = {41},
  pages   = {6259--6293},
  doi     = {10.1039/C2CS35070J}
}

@article{taatjes:2013,
  author  = {Taatjes, Craig A.},
  title   = {Criegee intermediates: What direct production and
                  detection can teach us about reactions of carbonyl
                  oxides},
  journal = arpc,
  year    = {2013},
  volume  = {64},
  pages   = {387--410},
  doi     = {10.1146/annurev-physchem-040412-110147}
}

@article{berndt:2012,
  author  = {Berndt, Torsten and Jokinen, Tuija and Sipilä, Mikko and
                  Mauldin, Roy L. and Herrmann, Hartmut and Stratmann,
                  Frank and Junninen, Heikki and Kulmala, Markku},
  title   = {Gas-Phase Ozonolysis of Selected Olefins: The Yield of
                  Stabilized Criegee Intermediate and the Reactivity
                  toward SO$_2$},
  journal = sci,
  year    = {2012},
  volume  = {335},
  pages   = {204--208},
  doi     = {10.1126/science.1213229}
}

\end{document}